\documentclass[preprint,amsmath,amssymb,aps]{revtex4-2}

\usepackage{natbib}

\usepackage{graphicx}
\usepackage{dcolumn}
\usepackage{bm}
\usepackage{multirow} 
\usepackage[normalem]{ulem}
\usepackage[labelfont=bf]{caption}
\usepackage{subcaption}
\usepackage[dvipsnames]{xcolor}

\begin{document}

\preprint{arXiv}

\title{Drainage front width in a three-dimensional random porous medium under gravitational and capillary effects}

\author{Paula Reis$^{1}$} 
\email{paula.reis@mn.uio.no}
\author{Knut J{\o}rgen M{\aa}l{\o}y $^{1,2}$}
\email{maloy@fys.uio.no}
\affiliation{$^1$PoreLab, The Njord Centre, Department of Physics, University of Oslo, P.O. Box 1048 Blindern, 0316 Oslo, Norway}
\affiliation{$^2$PoreLab, Department of Geoscience and Petroleum, Norwegian University of Science and Technology, 7031 Trondheim, Norway}

\date{\today}

\begin{abstract}

A theoretical approach to estimating stable drainage front widths in three-dimensional random porous media under gravitational and capillary effects is presented. Based on the frontier of the infinite cluster in gradient percolation, we propose an expression for the 3D front width dependent on the pore-network topology, the distribution of capillary pressure thresholds for the pore throats, the stabilizing capillary pressure gradient, the average pore size, and the correlation length critical exponent from percolation in three dimensions. Theoretical predictions are successfully compared to numerical results obtained with a bond invasion-percolation model for a wide range of drainage flow parameters. 

\end{abstract}

\maketitle

\section{Introduction}
\label{sec:intro}

The study of drainage in porous media is relevant to several subsurface transport processes and industrial applications \cite{juanes2010footprint,aryana2012experiments,datta2013drainage,hoogland2016drainage,hoogland2016drainageb,hoogland2017capillary,zacharoudiou2018impact}. It describes the displacement of a wetting fluid from the pore space by an immiscible non-wetting fluid, as their interface moves along a path of least resistance. At the pore scale, this movement is subject to the local capillary pressure value $P$ (defined as $P=P_{nw}-P_w$, where $P_{nw}$ is the pressure in the non-wetting phase, and $P_w$ is the pressure in the wetting phase) overcoming the capillary pressure threshold $P_t$ of a pore body or pore throat. Depending on the spatial distribution of $P_t$ values across the porous medium and variations of $P$ along the invasion front, diverse drainage patterns can arise, ranging from compact to ramified and fractal \cite{maaloy1985viscous,lenormand1985invasion,homsy1987viscous,lenormand1989capillary,lenormand1990liquids,birovljev1991gravity,maaloy1992dynamics,vincent2022stable}. Variations in $P$ along the interface are commonly related to different pressure gradients developed within each phase during drainage, often caused by gravitational and viscous forces. Values of $P_t$ are inversely proportional to the sizes of pore bodies and throats, and their spatial distribution in the medium can be homogeneous or heterogeneous \cite{meakin1991invasion,mani1999effect,vincent2022stable}.

Consider a porous medium where $P_t$ values for the pore throats are randomly drawn from an arbitrary distribution $N(P_t)$. In a drainage process with constant capillary pressure along the front, the non-wetting fluid invasion of such porous medium is characterized by capillary fingers \cite{lenormand1989capillary}, described well by the Invasion Percolation (IP) method \cite{wilkinson1983invasion}. Similarities between IP and ordinary percolation allow the theoretical estimation of relevant quantities related to drainage processes, such as fluid saturations and trapped cluster size distributions \cite{wilkinson1983invasion, lenormand1985invasion,lenormand1989capillary}. By associating each pore throat in a porous medium with a bond in a lattice, the minimum capillary pressure for the percolation of a porous medium, $P_{crit}$, can also be estimated \cite{birovljev1991gravity,auradou1999competition,moura2015impact}. For this, each pore throat capillary pressure threshold $P_t$ is associated with a bond occupation probability $p$, and $P_{crit}$ is the capillary pressure value that satisfies:

\begin{equation}
\label{eq:pc_crit}
    p_c=\int_{-\infty}^{P_{crit}} N(P_t) \,dP_t 
\end{equation}

\noindent{where $p_c$ is the bond percolation threshold for the lattice representing the porous medium.} 

In the same isotropic random porous medium, gradients in the capillary pressure along the interface may stabilize or destabilize the invasion front, significantly affecting the efficiency of the drainage process \cite{maaloy1985viscous,birovljev1991gravity,meakin1992invasion,meakin1992gradient,frette1992buoyancy,birovljev1995migration,wagner1997fragmentation,auradou1999competition,meheust2002interface}. Positive capillary pressure gradients in the direction of the flow lead to the unbounded growth of thin drainage fingers, which bypass most of the defending phase \cite{frette1992buoyancy}. Negative gradients limit the invasion front width to finite values, resulting in compact drainage patterns and a more efficient displacement of the defending phase. In this work, we focus our analysis on the drainage of a random 3D porous medium under gravitational and capillary effects, where a less dense non-wetting fluid displaces a denser wetting fluid from the top. Under these conditions, the gradient of capillary pressures at the interface is $G=-\Delta\rho g$, where $\Delta\rho$ is the fluid density difference and $g$ is the gravitational acceleration. In particular, we investigate the dependence of the stable drainage front width, $\eta_{3D}$, on the 3D porous medium structure and $G$.

A common approach to investigate gradient stabilized drainage in porous media is to establish a parallel between the fluids' interface and the frontier of the infinite cluster in the gradient-percolation problem \cite{sapoval1985fractal, rosso1986gradient, gouyet1988fractal}.  For 2D systems, \citet{sapoval1985fractal} demonstrated that this frontier takes place in a critical region, exhibiting a width $\eta/l$ of the order of the correlation length $\xi$, where $l$ is the length between two lattice sites. As further explained in Section~\ref{sec:theo}, the analogy between $\eta/l$ and $\xi$ leads to a drainage interface scaling of the type:

\begin{equation}
    \eta/a \propto F^{-\nu/(1+\nu)}
    \label{eq:f_scaling_2D}
\end{equation}

\noindent{where $a$ is the average pore length, $\nu=4/3$ is the percolation critical exponent for the correlation length in 2D, and $F$ is the dimensionless fluctuation number, $F=-N(P_{crit})Ga$, defined in \cite{auradou1999competition}.}

The theoretical prediction in Eq.~\ref{eq:f_scaling_2D} has been verified by numerous studies of drainage in 2D porous media, both numerically and experimentally  \cite{birovljev1991gravity,meakin1992gradient,auradou1999competition,meakin2000invasion,zhang2000spreading,meheust2002interface,lovoll2005competition}. In 3D porous media, however, the same interface scaling is not expected to be valid. By extending the gradient-percolation problem to 3D lattices, \citet{gouyet1988fractal} demonstrated that the infinite cluster frontier is no longer confined to a critical region but rather contains a critical region, termed the front tail. Only within the length of the front tail, $\eta_{t}$, does the front present a fractal structure and obey scaling laws obtained from percolation theory. In the context of drainage, \citet{chaouche1994invasion} performed experiments with 3D porous media consisting of packed glass beads and verified that $\eta_{t}\propto Bo^{-\nu/(1+\nu)}$, where $Bo$ is the Bond number. Similar scaling laws have been proposed by \citet{wilkinson1984percolation} for the maximum length of trapped wetting-fluid clusters, $L_{max}$, and for the residual wetting-fluid saturation, $S_{res}$.
While both represent relevant drainage parameters, they are insufficient to disclose the full extent of the region spanned by the front.

In the next section, we build on the theoretical approach presented by \citet{gouyet1988fractal} to propose an equation for $\eta_{3D}$ in gradient stabilized drainage in three-dimensional random porous media. The prediction is later verified with a modified IP model incorporating capillary and gravitational effects in Sec.~\ref{sec:IP}. 

\section{Theory}
\label{sec:theo}

In two dimensions, gradient percolation describes the behavior of $N\times N$ lattices where the site or bond occupation probability $p$ varies monotonically along a direction $z$. Let us assume a percolation problem in a square lattice, where $p$ varies from $p(z=0)=1$ to $p(z=N)=0$. In this lattice, an infinite cluster exists, containing the occupied sites or bonds connected to the boundary where $p=1$. The infinite cluster presents a fractal frontier of width $\eta$, centered at $z_c$, where $p(z=z_c)=p_c$. Considering that clusters formed in the frontier region should be of the order of the correlation length $\xi \propto |p-p_c|^{-\nu}$ (where $\nu$ is the percolation critical exponent for the correlation length, $\xi$) and bounded by the frontier width, \citet{sapoval1985fractal} reached the following scaling for $\eta$:

\begin{equation}
\label{eq:sapoval}
    \eta \propto \left|\left(\frac{\partial p}{\partial z} \right)_{z=z_c}\right|^{-\nu/(1+\nu)}
\end{equation}

Analogous theoretical considerations can be used for the width $\eta$ of a stable drainage front in 2D porous media, where $\nabla P = G$ is a linear capillary pressure gradient in the flow direction, $z$. As proposed by \citet{birovljev1991gravity}, at some point $z_c$ in the drainage front, the capillary pressure is $P_{crit}$, where the probability of invasion of a pore throat is $p_c$ (see Eq. ~\ref{eq:pc_crit}). Based on that, the probability of invasion $p$ of a pore throat in the front at an arbitrary position $z=z_c+\Delta z$ can be calculated as:

\begin{equation}
\label{eq:occ_prob}
    p=p_c+\int_{P_{crit}}^{P_{crit}+G\Delta z} N(P_t) \,dP_t 
\end{equation}

Approximating the solution of Eq.~\ref{eq:occ_prob} by the lowest order term of the Taylor expansion of $N(P_t)$, we get $p\approx p_c + N(P_{crit})G\Delta z$. Therefore, the pore throats in the drainage front display a linear gradient in invasion probability near the critical value $p_c$, in the same way as the bonds in the frontier of the infinite cluster in gradient percolation. We can assume then that the drainage front width can be related to the correlation length as $\eta=\xi a$, where $a$ is the average pore size. Substituting $|\Delta z|$ by $\eta$ gives $|p-p_c|=N(P_{crit})G\xi a \propto \xi^{-1/\nu}$, which is equivalent to the scaling relation for the front width introduced by Eq.~\ref{eq:f_scaling_2D}.

A fundamental aspect for the validity of the scaling presented in Eq.~\ref{eq:sapoval} is the percolation threshold of the lattice of interest \cite{rosso1986gradient}. In 2D lattices, $p_c$ values often exceed 0.5. Considering that occupied sites or bonds in a lattice represent the non-wetting phase in pore bodies and throats during drainage, $p_c>0.5$ implies that only one phase can percolate at a time. As the minimum capillary pressure required for the percolation of the non-wetting phase is achieved (see Eq.~\ref{eq:pc_crit}), the drainage of a large fraction of pores is allowed, which traps the wetting phase in smaller clusters. For this reason, the drainage front does not extend over a large range of $p$, being restricted to the critical region around $p_c$, where Eq.~\ref{eq:sapoval} is valid.

On the contrary, three-dimensional lattices can exhibit $p_c$ values significantly lower than 0.5. This means that a phase can percolate the lattice-equivalent porous medium at a relatively low saturation, without restricting a second phase to isolated clusters. More precisely, there is a range of occupation probabilities $p_c\leq p\leq 1-p_c$ over which drainage fronts can be extended, as both phases reach the percolation threshold. As a consequence, the capillary pressure values at the 3D drainage front vary approximately from $P_{crit}$ at its tip, where the invading phase saturation is low, to $P_{res}$ at its end, where the invading phase saturation is high enough to trap the defending phase at the residual saturation $S_{res}$.  Here, $P_{res}$ is given by:

\begin{equation}
\label{eq:pc_trap}
    1-p_c=\int_{-\infty}^{P_{res}} N(P_t) \,dP_t 
\end{equation} 

In this scenario, as identified by \citet{rosso1986gradient}, the stretch of the front beyond the vicinity of $p_c$ makes the front width scaling presented by \citet{sapoval1985fractal} no longer valid. Instead, \citet{gouyet1988fractal} demonstrated that only a tail of the front, $\eta_{t}$, spanning the critical region near $p_c$, exhibits a fractal structure. Using a simple cubic lattice and different linear gradients $\nabla p$, they measured the subregion of the infinite cluster frontier limited by $p\leq p_c$ and verified that $\eta_{t} \propto \left|\nabla p \right|^{-\nu/(1+\nu)}$. Following a similar reasoning, we can assume that 3D stable drainage fronts also present a fractal tip region, $\eta_t$, with capillary pressure values $P\leq P_{crit}$, which scales as $\eta_t/a\propto F^{-\nu/(1+\nu)}$.

Beyond what is proposed by \citet{gouyet1988fractal}, we suggest that within 3D stable drainage fronts in porous media, a second fractal region near $p=1-p_c$ exists, where capillary pressure values are $P\geq P_{res}$. Analogous to the tip of the drainage front, where the non-wetting phase is near its percolation threshold, the end of the front should also exhibit a critical behavior, as the wetting phase is about to exit its percolation range. The width of this fractal region, termed here $\eta_r$, should be of the order of the maximum length of the trapped wetting-phase clusters $L_{max}$ \cite{sapoval1985fractal,gouyet1988fractal}, and scale as $\eta_r/a\propto F_r^{-\nu/(1+\nu)}$ (where $F_r=-N(P_{res})Ga$ is a modified Fluctuation number). 

Similarly, \citet{wilkinson1984percolation} suggested that $L_{max}\propto Bo^{-\nu/(1+\nu)}$, but no link between this scaling and the front width was established. Rather, a transition zone $h$ was defined in that study, bounded by the planes where the invading phase occupation is $p_c$ and $1-p_c$, which was considered to scale as $h\propto Bo^{-1}$. We propose that this transition zone can be directly calculated as $h=(P_{res}-P_{crit})|G|^{-1}$, and corresponds to the distance between the two fractal regions, $\eta_t$ and $\eta_r$. In this way, we can write a function for the full 3D front length equivalent to:

\begin{equation}
    \label{eq:FF_3D_front}
    \eta_{3D}=(P_{res}-P_{crit})|G|^{-1}+a\mathcal{C}\left(F^{-\nu/(1+\nu)}+F_r^{-\nu/(1+\nu)}\right)
\end{equation}

\noindent{where $\mathcal{C}$ is of the order of unity, and is related to $\xi_0$ in $\xi=\xi_0|p-p_c|^{-\nu}$, the correlation length scaling in percolation \cite{stauffer1985introduction}. In 3D, $\nu=0.88$}. 

\section{Invasion Percolation Modeling}
\label{sec:IP}

To test the theoretical assumptions presented in Sec. \ref{sec:theo}, an Invasion Percolation (IP) model is used. Proposed in the 1980s by \citet{wilkinson1983invasion}, IP modeling was conceived to represent quasi-static porous media displacement flows dominated by capillary forces. Since then, it has been intensively adopted in the porous-media literature, especially to represent slow drainage flows \cite{maaloy1992dynamics,furuberg1996intermittent,lovoll2005competition,masson2014fast,vincent2022stable,reis2023simplified,khobaib2025gravity}.
In IP models, a network of sites connected by bonds represents a porous medium of pore bodies connected by pore throats. To each site or bond, a capillary pressure threshold $P_t$ for invasion is assigned, which controls its likelihood of invasion in the pore network. Drainage is modeled as a bond-invasion percolation process, where bonds with low $P_t$ are preferentially invaded. In imbibition, a site-invasion percolation process is adopted, where sites with high $P_t$ are more likely to be invaded.

Briefly, our IP simulations represent drainage flows and start with a pore network fully occupied by the wetting phase. From sites defined as the pore-network inlet, the occupation of the non-wetting phase evolves as a growing cluster that sequentially incorporates the ``easiest" available bond at its perimeter. During this process, wetting-phase clusters completely surrounded by the invading phase are considered trapped. Therefore, available bonds belong only to wetting-phase clusters connected to the network outlet. The bond-invasion percolation progresses until the breakthrough -- when the invading phase reaches the network outlet -- and the drainage simulations end. For more details about the Invasion Percolation method, we refer to \citet{wilkinson1983invasion}.

\subsection{Stable drainage under gravitational effects}
\label{sec:grav_eff}

Within drainage flows, we focus on the stable scenario in which a denser wetting fluid is slowly displaced by a less dense non-wetting fluid from the top. Thus, gravitational effects are included in the model by considering that the capillary pressure in the pore throats at the drainage front varies linearly, according to Eq. \ref{eq:Pc_z} \cite{wilkinson1984percolation,auradou1999competition}. 

\begin{equation}
    P(z)=P_0-\Delta\rho g z
    \label{eq:Pc_z}
\end{equation}

\noindent{where $P_0$ is a reference value of capillary pressure at $z=0$, and the $z$ axis points in the direction of the gravitational field.}

As a consequence, our bond invasion-percolation model is controlled by both the capillary pressure threshold values and the position of the bond along the invasion front. In this way, at each invasion step, the ``easiest" available bond at the invasion cluster perimeter corresponds to the one with the lowest value of $P_t^*=P_t-P(z)$.

\subsection{Networks representing porous media}
\label{sec:nets}

Networks with different topologies, capillary pressure threshold distributions $N(P_t)$, and pore sizes $a$ are used to represent porous media in the IP drainage simulations presented in this work. With this, we aim to verify that the theoretical predictions presented in Sec. \ref{sec:theo} are valid for pore-networks with variable geometrical features.

\subsubsection{Network Topology}

Two regular 3D structures are considered: a simple cubic structure, in which each site has six nearest neighbors, and a diamond cubic structure, in which each site has four nearest neighbors. Similar values of pore-network coordination numbers have been reported in the literature for relevant naturally occurring porous media, such as sandstones \cite{lindquist1999investigating,ioannidis2000geometry,vasilyev2012effect}. Due to their difference in connectivity, the evaluated structures display significantly different bond percolation thresholds. simple-cubic networks present a bond percolation threshold $p_c\approx0.25$, while this value is $p_c\approx0.39$ for diamond-cubic networks. Based on that, important differences during the drainage of these structures are expected. With a lower $p_c$, porous media corresponding to a simple cubic lattice can be percolated at a lower critical capillary pressure $P_{crit}$, present a wider range of occupations where both phases percolate ($p_c\leq p\leq 1-p_c$), and require higher capillary pressure values to trap the wetting phase, $P_{res}$.

\subsubsection{Capillary-pressure-threshold distributions}

\begin{figure}[ht!]
     \centering
     
     \begin{subfigure}[t]{0.48\textwidth}
         \centering
         \includegraphics[width=1\textwidth]{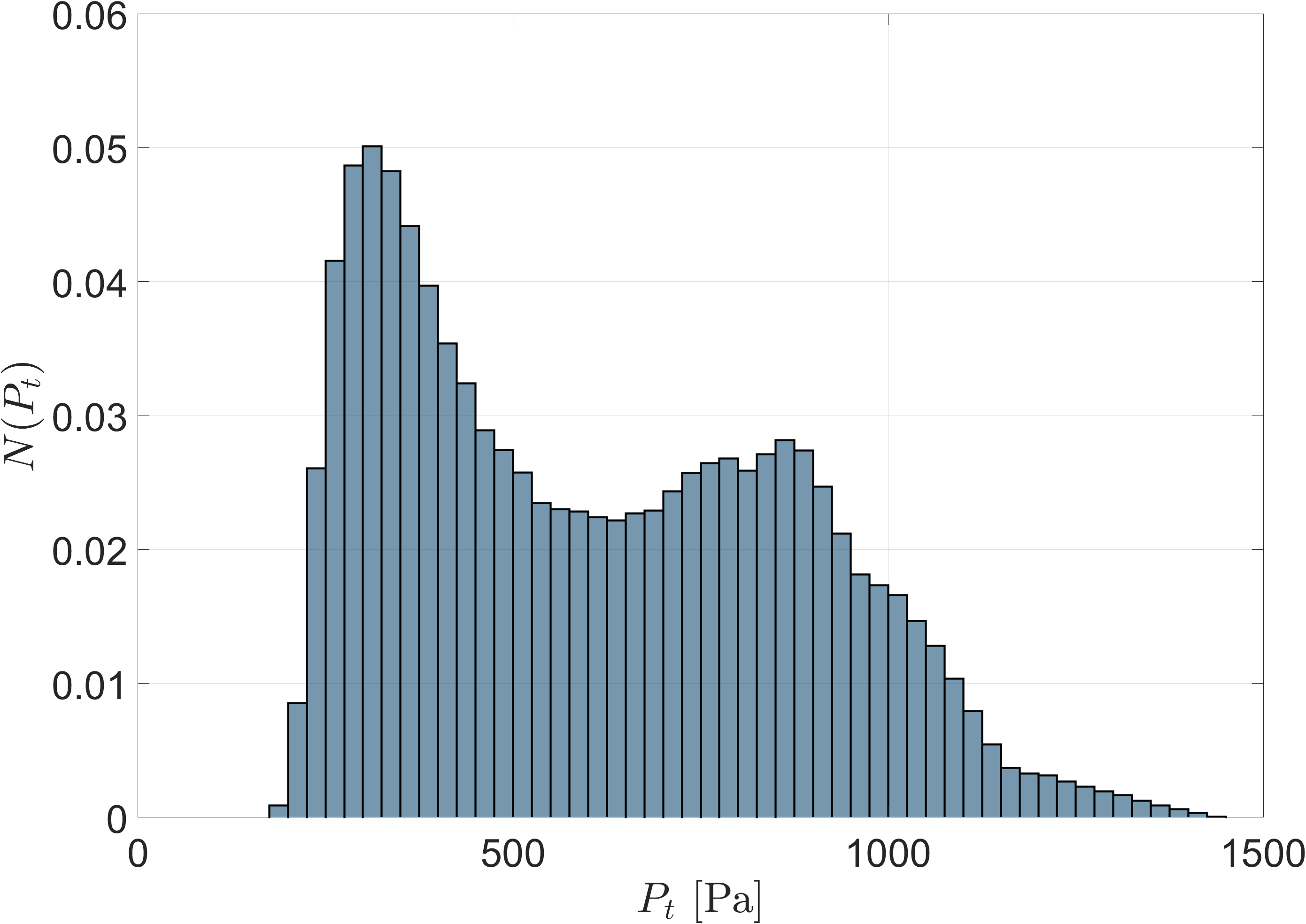}
         \caption{Non-Uniform $N(P_t)$}
         \label{fig:MouraNPt}
     \end{subfigure}
     \hfill
     \begin{subfigure}[t]{0.48\textwidth}
         \centering
         \includegraphics[width=1\textwidth]{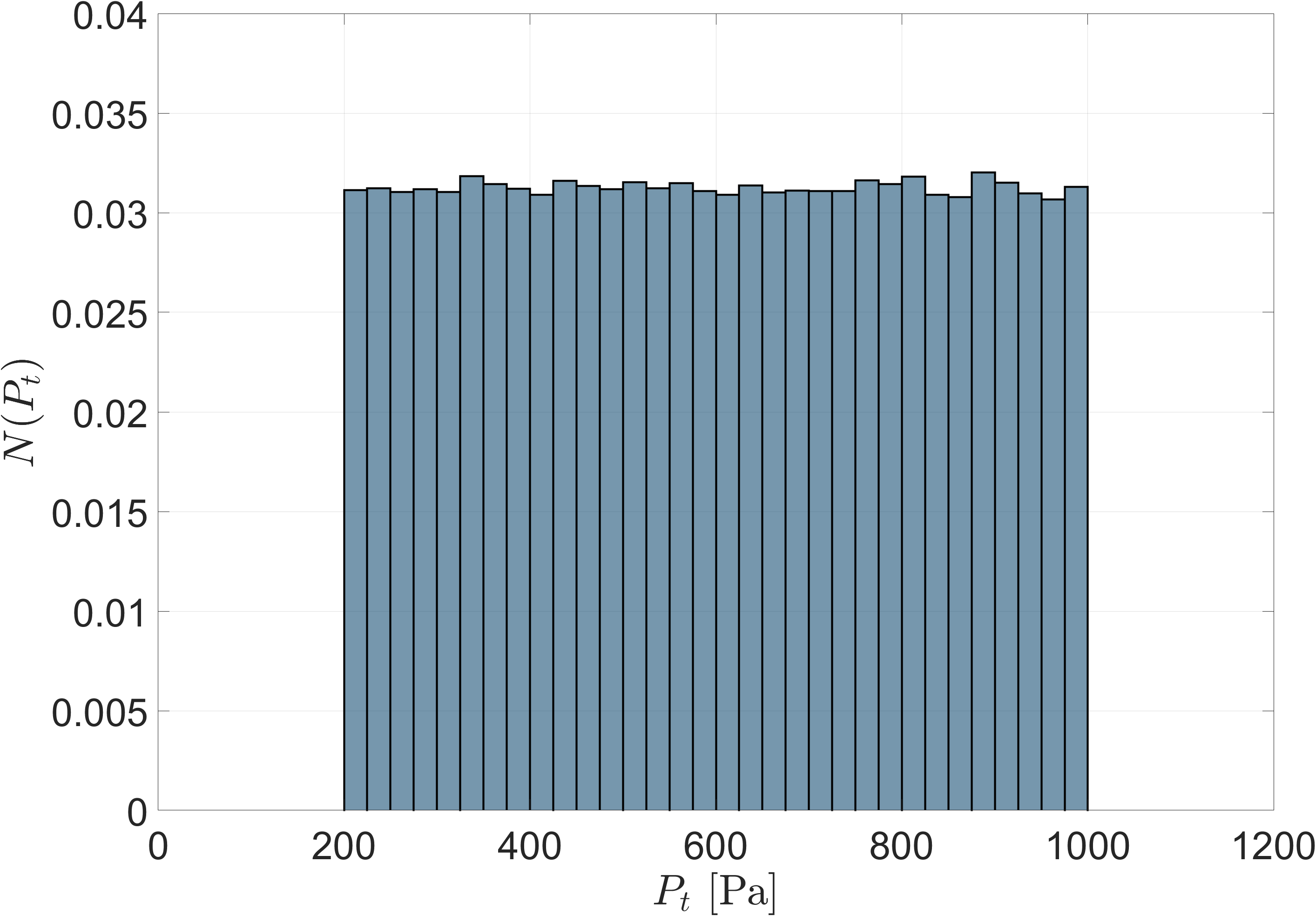}
         \caption{Uniform $N(P_t)$}
         \label{fig:UniformNPt}
     \end{subfigure}
     \hfill
     
        \caption{Histogram of the $P_t$ distributions used to generate the pore networks, with a bin width equivalent to 25 Pa.}
        \label{fig:NPt}
\end{figure}

Both a uniform and a non-uniform $N(P_t)$ distribution are used in the simulations, as shown in Fig. \ref{fig:NPt}. The distribution presented in Fig. \ref{fig:MouraNPt} is based on values reported in \citet{moura2019connectivity}, while the one presented in Fig. \ref{fig:UniformNPt} is simply a uniform distribution within the range $P_t\in[200,1000]$ Pa. From these distributions, values are assigned to each bond in the networks randomly, without establishing a spatial correlation between $P_t$ values.

In Table \ref{tab:Pcrit}, the critical capillary pressures for percolation of the non-wetting phase (see Eq. \ref{eq:pc_crit}) and trapping of the wetting phase (see Eq. \ref{eq:pc_trap}) are shown. Also, the values of $N(P_{crit})$ and $N(P_{res})$ -- which are used to calculate the Fluctuation numbers at the critical capillary pressures -- are calculated. For the uniform capillary pressure threshold distribution, we have a constant probability of occurrence for any $P_t$, equal to $N(P_t)=1/W_t$, where $W_t$ is the width of the distribution. For the non-uniform case, we calculate $N(P_t)$ with the numerical approximation of the derivative of the cumulative distribution function of $P_t$ at $P_{crit}$ and $P_{res}$, similarly to the procedure presented in \citet{khobaib2025gravity}.

\begin{table}[b]
\caption {\label{tab:Pcrit} Critical capillary pressure values and their probability in $N(P_t)$} 
\begin{center}    
\begin{tabular}{ c c c c c   }
 \hline
 \hline
 \multicolumn{1}{l}{}&
 \multicolumn{2}{c}{Diamond Cubic}&
 \multicolumn{2}{c}{Simple Cubic}\\
     & Non-Uniform & Uniform & Non-Uniform  & Uniform \\
 \hline
$P_{crit}$ [Pa] & 463.3 & 511.1 & 364.8 & 400 \\
$P_{res}$ [Pa] & 693.5 & 688.9 & 830.8 & 800 \\
 $N(P_{crit})$ [Pa$^{-1}$]  & $1.13\times10^{-3}$& $1.25\times10^{-3}$ & $1.77\times10^{-3}$ & $1.25\times10^{-3}$ \\
$N(P_{res})$ [Pa$^{-1}$]  & $9.34\times10^{-4}$ & $1.25\times10^{-3}$& $1.06\times10^{-3}$ & $1.25\times10^{-3}$ \\
 \hline
\hline
\end{tabular}
\end{center}
\end{table}

\subsubsection{Average pore size}

Three different values of pore size $a=[0.1,0.5,1]$ cm are used in the drainage simulations. In this way, a wider range of invasion front widths, as shown in the next section, could be covered by this study. It is important to notice that this size is not equivalent to the opening of the constrictions -- or pore throats -- in the media. While the constrictions radii are related to the $N(P_t)$, shown in Fig. \ref{fig:NPt}, $a$ is related to the average distance between pore bodies projected in the direction $z$, which is aligned with the gravitational field $g$. Hence, this quantity can be linked to the length of pores.

\subsubsection{Pore-network size and boundary conditions}

For each pore-network topology, a different network size is used in the simulations. diamond-cubic networks displayed a $n_x\times n_y \times n_z$ of $100\times100\times200$ sites, where $n_x$, $n_y$, and $n_z$ are the number of sites in the $x$, $y$ and $z$ directions, respectively, and $z$ is aligned with the gravitational field. simple-cubic networks had half the number of sites in each direction, namely $50\times50\times100$, as their coordination number is higher and their bond percolation threshold is lower, which leads to more computationally expensive simulations.

\begin{figure}[ht!]
     \centering 
        \includegraphics[width=0.25\textwidth]{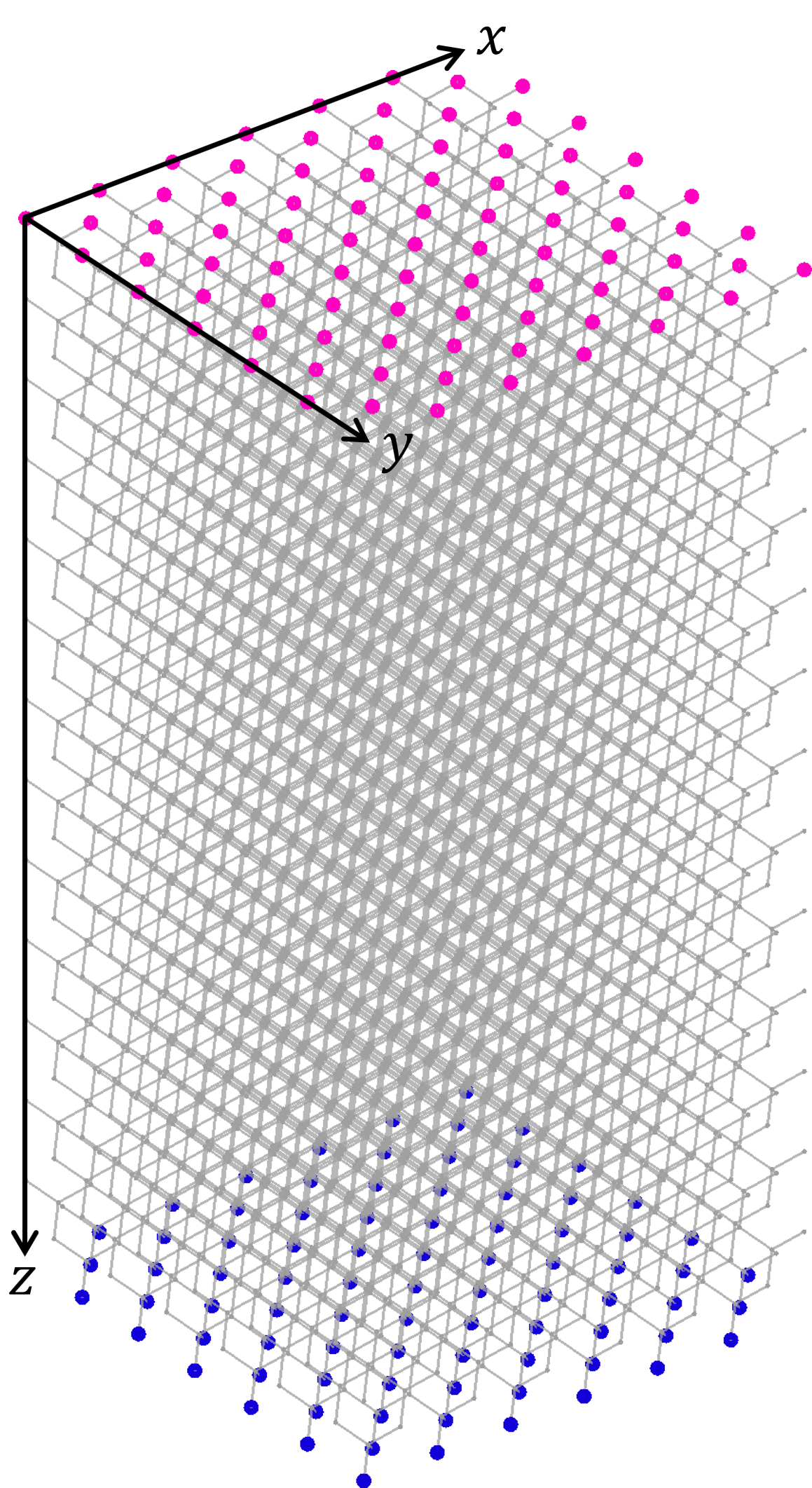}
        \caption{Illustration of a diamond-cubic network with inlet sites highlighted in pink and outlet nodes highlighted in blue.}
        \label{fig:nodes_in_out}
\end{figure}

As for the boundary conditions, all sites at $z=0$ are assigned to the network inlet, while all sites at $z=an_z$ are assigned to the network outlet, as shown in Fig. \ref{fig:nodes_in_out}. The network sides -- corresponding to $x=0$, $x=an_x$, $y=0$, and $y=an_y$ -- are considered closed.

\subsection{Imposed gradient of capillary pressure at the drainage fronts}
\label{sec:delta_rho}

To obtain drainage fronts under different capillary pressure gradients $G$, we vary the density difference between the phases in the IP drainage simulations. For each type of pore network used in this study, a different range of $\Delta\rho$ is defined.

At the low-density-contrast end of the range, drainage fronts are wide, and values of $\Delta\rho$ are chosen so that the obtained front fits the total size of the pore network. For this, some tests are conducted for each type of network until a minimum suitable value of $\Delta\rho$ is found. At the other end of the range, drainage fronts become narrow, and we seek to avoid nearly flat fronts.
Thus, we set the maximum value of $\Delta\rho$ when front widths are $\approx5a$. Therefore, in the results presented in Sec. \ref{sec:results}, we include drainage cases with stable front widths belonging approximately to the interval $5a<\eta_{3D}<n_za$.

\section{Results}
\label{sec:results}

In this section, we use the bond invasion-percolation method detailed in Sec. \ref{sec:IP} to simulate multiple slow drainage scenarios under stabilizing gravitational effects. The observed front characteristics are compared to the theoretical predictions from Sec. \ref{sec:theo}. In particular, we aim to verify whether the estimate of the stable front width $\eta_{3D}$, proposed in Eq. \ref{eq:FF_3D_front}, is valid. 

\begin{figure}[ht!]
     \centering
     
     \begin{subfigure}[t]{0.25\textwidth}
         \centering
         \includegraphics[width=1\textwidth]{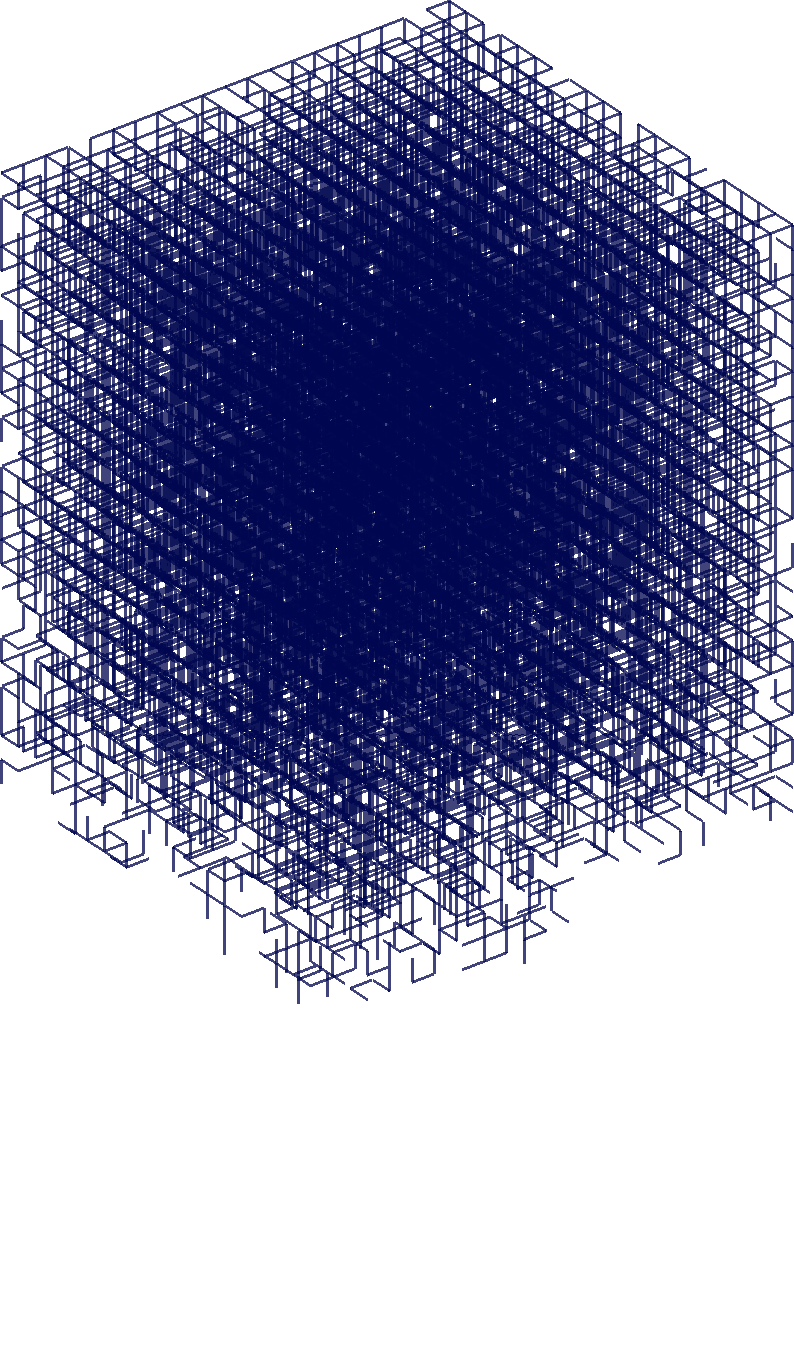}
         \caption{Bonds occupied by the invading phase}
         \label{fig:edges_inv}
     \end{subfigure}
     \begin{subfigure}[t]{0.25\textwidth}
         \centering
         \includegraphics[width=1\textwidth]{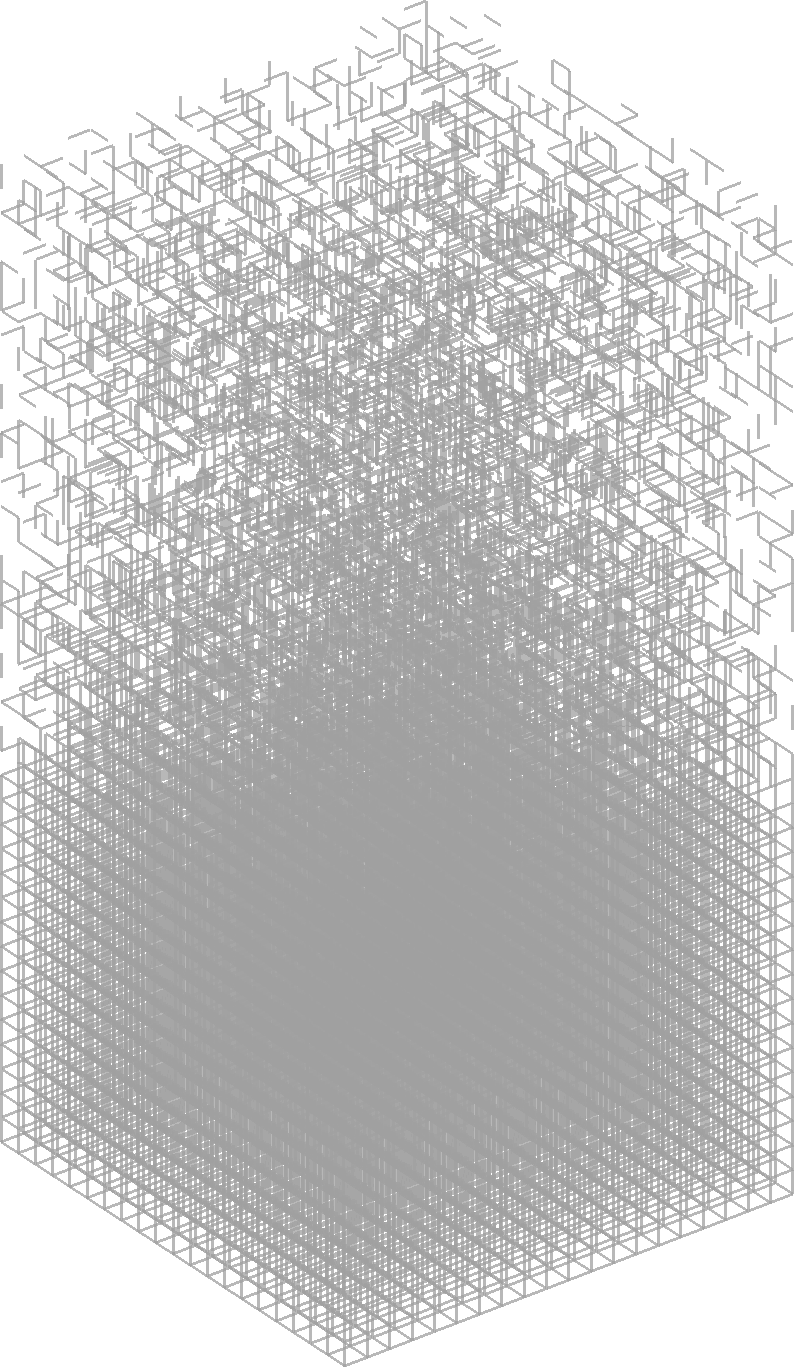}
         \caption{Bonds occupied by the defending phase}
         \label{fig:edges_rem}
     \end{subfigure}
     \begin{subfigure}[t]{0.25\textwidth}
         \centering
         \includegraphics[width=1\textwidth]{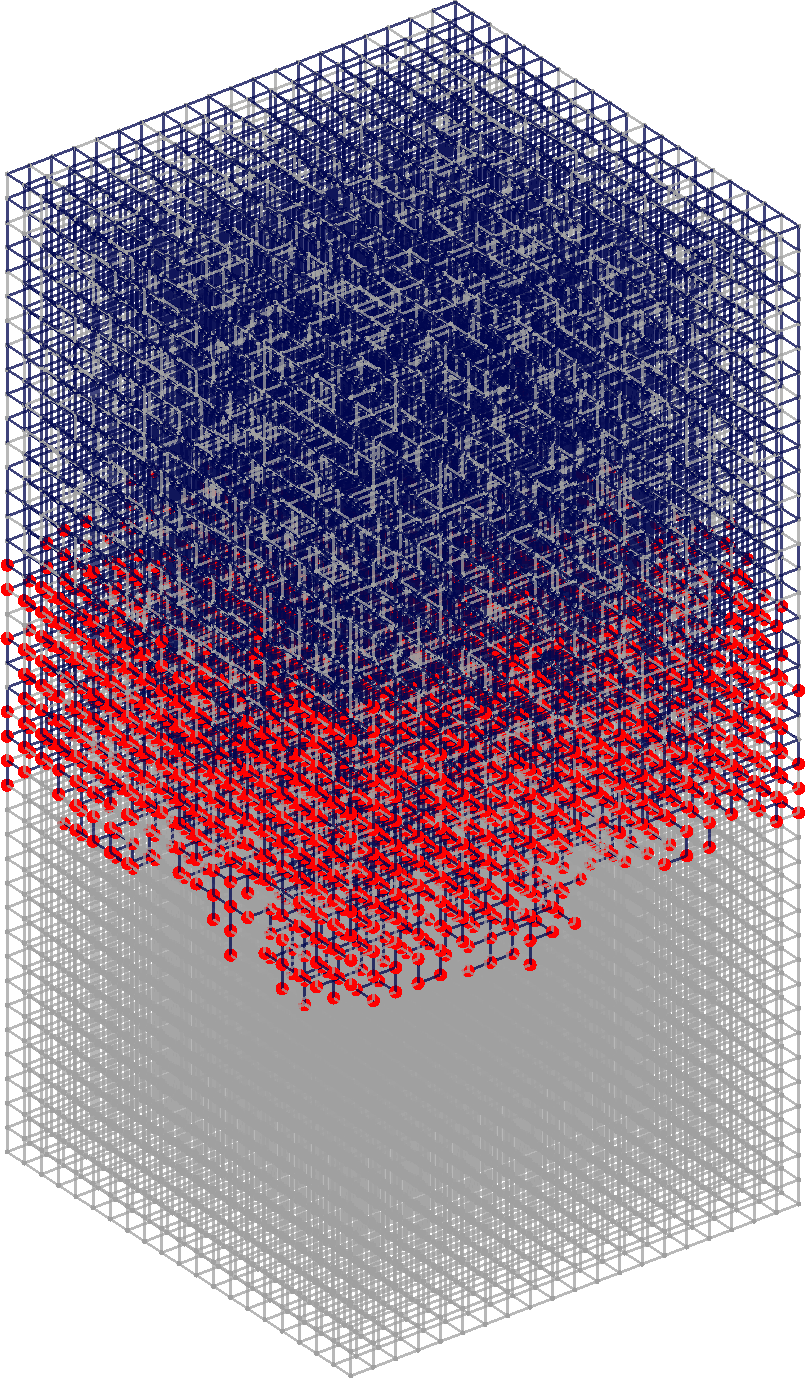}
         \caption{Sites at the invasion front}
         \label{fig:nodes_front}
     \end{subfigure}
     
        \caption{Simple-cubic network midway through drainage, with invaded bonds in blue, non-invaded bonds in gray, and sites belonging to the invasion front in red. In this example, $a=0.5$ cm and $\Delta\rho=1000$ kg/m$^3$}
        \label{fig:Graph_Front_Nodes}
\end{figure}

To calculate the stable front width from each drainage simulation, we keep track of the pore-network sites and bonds belonging to the invasion front, from the beginning of the non-wetting phase invasion to the breakthrough. Figure \ref{fig:Graph_Front_Nodes} illustrates a simple-cubic network, midway through drainage, with the bonds of the invading cluster highlighted in dark blue, the bonds of defending clusters in gray, and the sites belonging to the invasion front in red. 

Using the $x,y,z$ coordinates of the bonds connected to the front sites, as well as their $P_t$ values, we can estimate the transition zone width $h$ and the width of the critical zones $\eta_t$ and $\eta_r$ throughout each drainage simulation. Time-averaged values of these widths are presented in Sections \ref{sec:h}, \ref{sec:tail_crit}, and \ref{sec:tail_res}. Invasion fronts containing sites or bonds close to the inlet plane, namely $0\leq z\leq0.25an_z$, are not considered in the analyses, as they are prone to distortion by boundary effects. 

\subsection{Transition zone}
\label{sec:h}

As proposed by \citet{wilkinson1984percolation},  invasion-percolation stable fronts present a transition zone from the point where the occupation of the invading fluid is equivalent to the percolation threshold, to the point where the occupation of the defending fluid is at the percolation threshold. For any pore network with a defined $p_c$, the capillary pressure values corresponding to these percolation thresholds can be estimated with Eqs. \ref{eq:pc_crit} and \ref{eq:pc_trap}, as presented in Table \ref{tab:Pcrit}. Using these values, the limits of the transition zone within the stable fronts can be located, as explained next.

\begin{figure}[ht!]
     \centering
     
     \begin{subfigure}[t]{0.275\textwidth}
         \centering
         \includegraphics[width=1\textwidth]{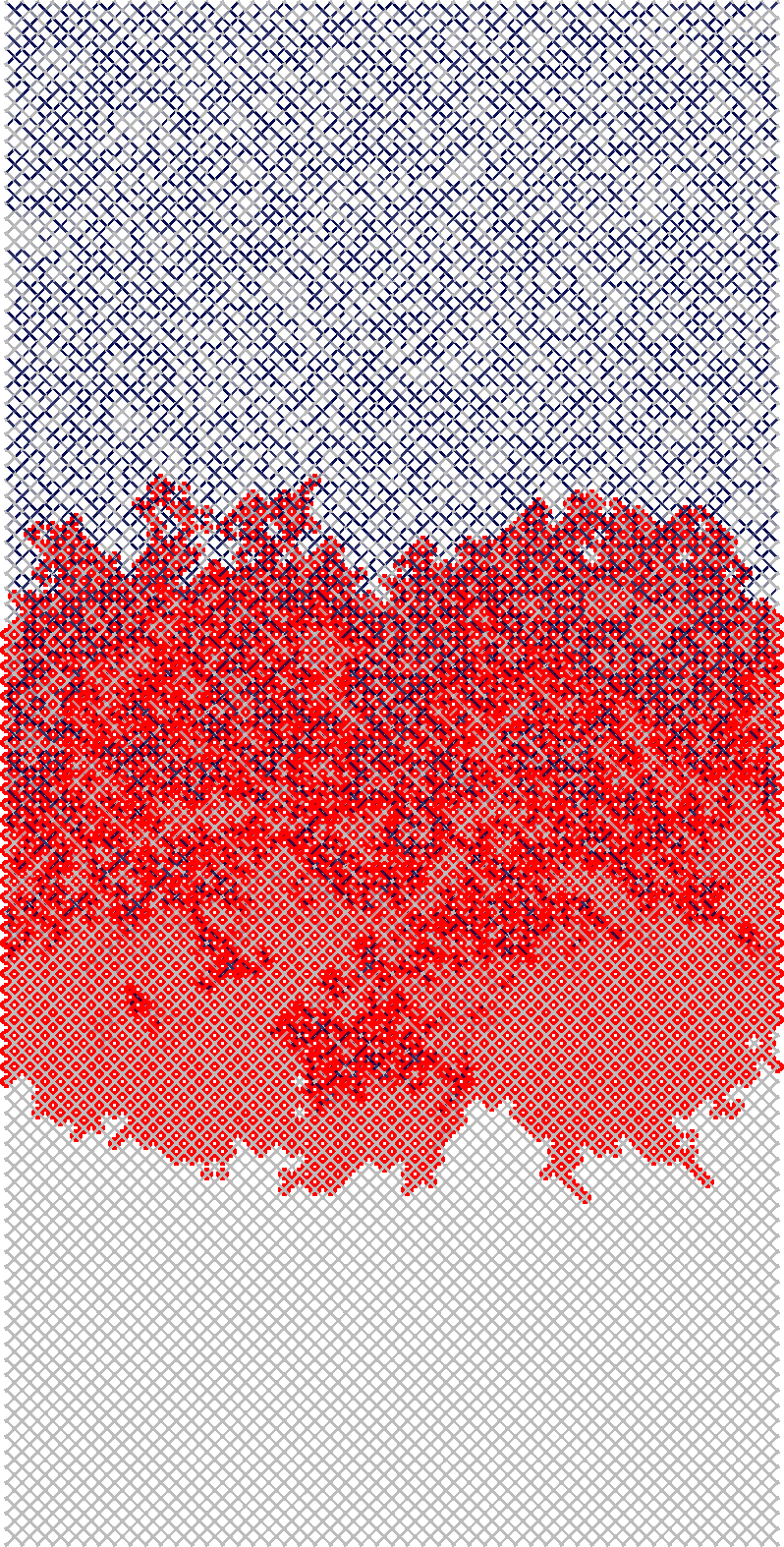}
         \caption{}
         \label{fig:nodes_front_example}
     \end{subfigure}
     \begin{subfigure}[t]{0.275\textwidth}
         \centering
         \includegraphics[width=1\textwidth]{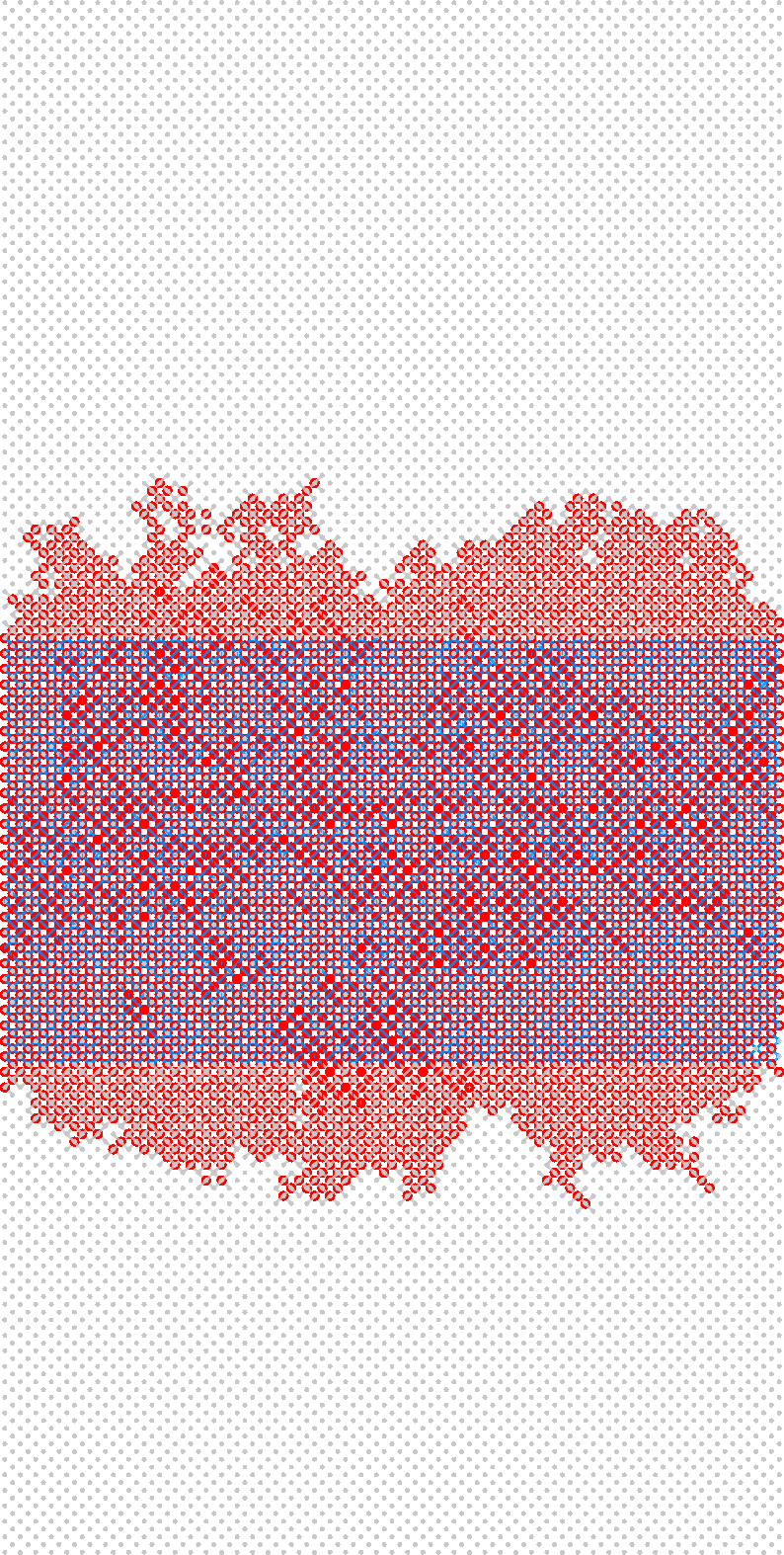}
         \caption{}
         \label{fig:nodes_center_example}
     \end{subfigure}
     \begin{subfigure}[t]{0.575\textwidth}
         \centering
         \includegraphics[width=1\textwidth]{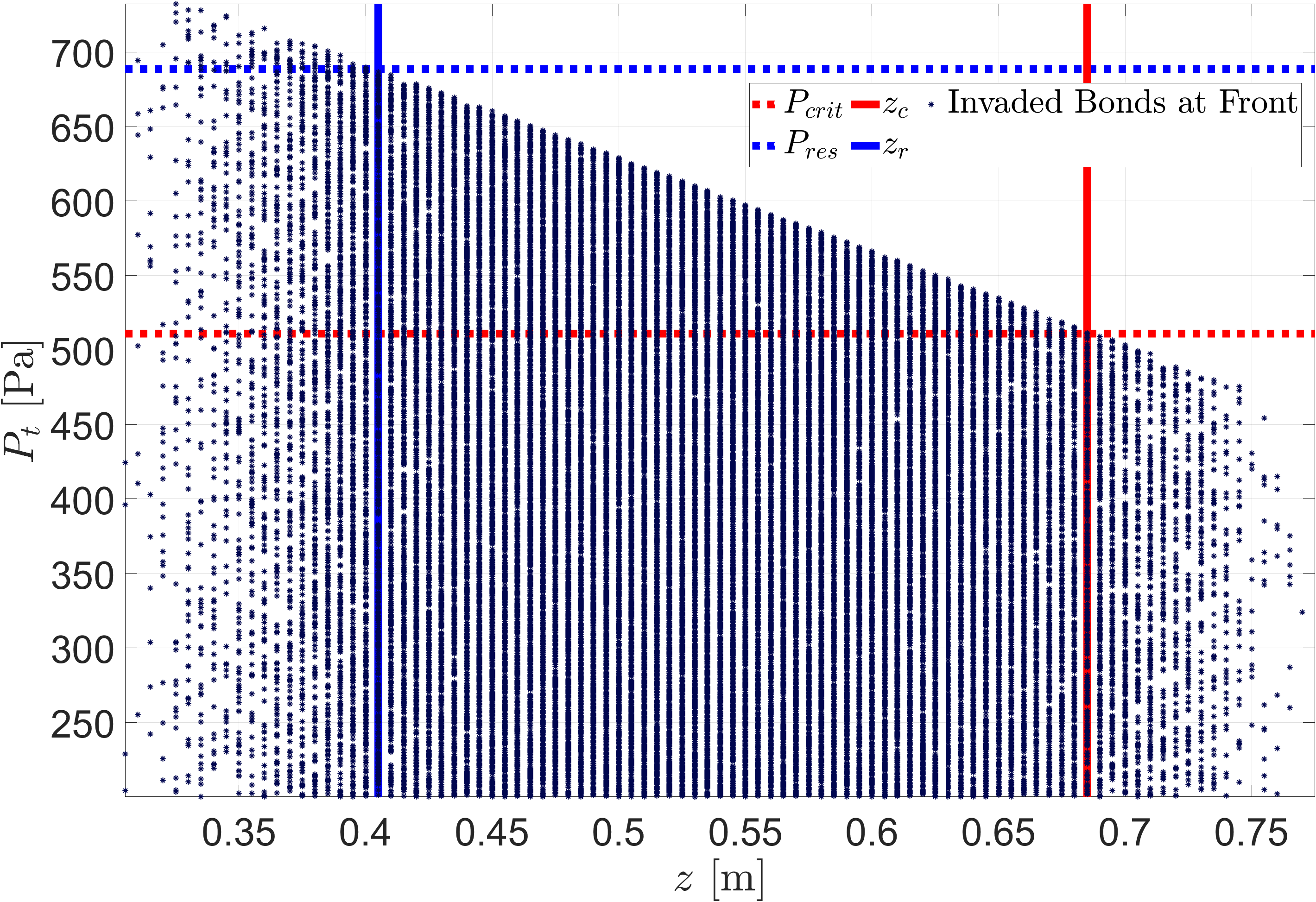}
         \caption{}
         \label{fig:pc_vs_z_invaded}
     \end{subfigure}
     \hfill
     
        \caption{Diamond-cubic network during drainage, viewed from the plane $x\times z$, with $a=0.5$ cm and $\Delta\rho=64$ kg/m$^3$. (a) Invaded bonds are shown in dark blue, non-invaded bonds in gray, and front sites in red. (b) The transition zone is shown in blue, and the front sites in red. Only bonds connected to the front are shown. (c) $P_t$ of the invaded bonds connected to the front $vs$ their position in the $z$ axis. $z_{c}$ and $z_{r}$ represent the lower and upper limits of the transition zone $h$, respectively.}
        \label{fig:front_center_example}
\end{figure}

Let us consider a diamond-cubic pore network, with $a=0.5$ cm and a uniform $N(P_t)$ from 200 Pa to 1000 Pa (see Fig. \ref{fig:UniformNPt}). For the non-wetting phase to percolate this medium, the critical capillary pressure is estimated to be $P_{crit}=511.1$ Pa. At $P_{res}=688.9$ Pa, we expect the occupation of the wetting phase to be near its critical value, when it becomes confined to trapped clusters. In Fig. \ref{fig:front_center_example}, we illustrate this network during drainage, when $\Delta\rho=64$ kg/m$^3$. As we aim to locate the points in the $z$ axis delimiting the transition zone $h$, a view from the plane $x \times z$ is chosen. In Fig. \ref{fig:nodes_front_example}, we show the invaded bonds in dark blue, the non-invaded bonds in gray, and the front sites in red. In \ref{fig:nodes_center_example}, the sites belonging to the front are still shown in red, and only bonds connected to the front are displayed. Bonds within the transition zone are highlighted in blue, and the other network elements are gray. Figure \ref{fig:pc_vs_z_invaded} presents the $P_t$ of all invaded bonds at the front, according to their position in the $z$ axis. It is clear that, as we move up in the front (in the negative direction of the $z$ axis), the maximum capillary threshold value allowed for invasion increases. This results from the linear gradient in capillary pressures along the front, $G=-\Delta\rho g$. At the front tip, $z\approx0.7$ m, the capillary pressure is relatively low, and only pore throats with $P_t$ up to $P_{crit}$ can be invaded. At the front end, near $z\approx0.35$ m, the local capillary pressure values are much larger, allowing the invasion of pore throats with $P_t\approx P_{res}$. Therefore, to define the limits of the transition zone $h$, we find the maximum value of $P_t$ among the invaded bonds at the invasion front, for each value of $z$ spanned by the front. At the points where these maxima correspond to $P_{crit}$ and $P_{res}$, we locate the limits of $h$. It is important to note that a well-behaved gradient in the maximum $P_t$ among invaded pore throats along the $z$ axis, as seen in Fig. \ref{fig:pc_vs_z_invaded}, may not be observed if the pore network cross section $(n_x \times n_y)$ is not large enough to contain a significant sample of $N(P_t)$. 

\begin{figure}[ht!]
     \centering
     
     \begin{subfigure}[t]{0.48\textwidth}
         \centering
         \includegraphics[width=1\textwidth]{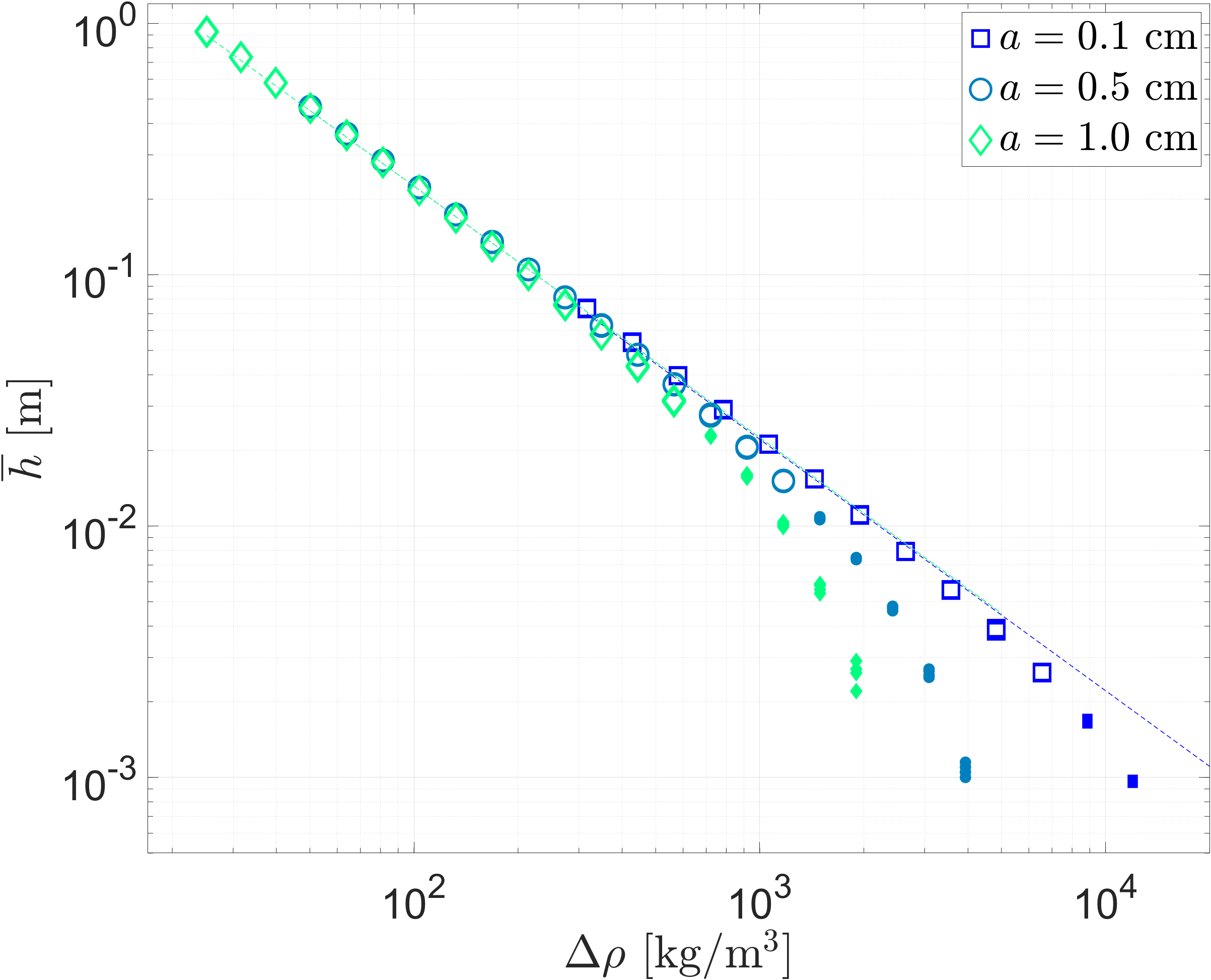}
         \caption{Diamond Lattice - Non-uniform $N(P_t)$}
         \label{fig:h_4_2}
     \end{subfigure}
     \hfill
     \begin{subfigure}[t]{0.48\textwidth}
         \centering
         \includegraphics[width=1\textwidth]{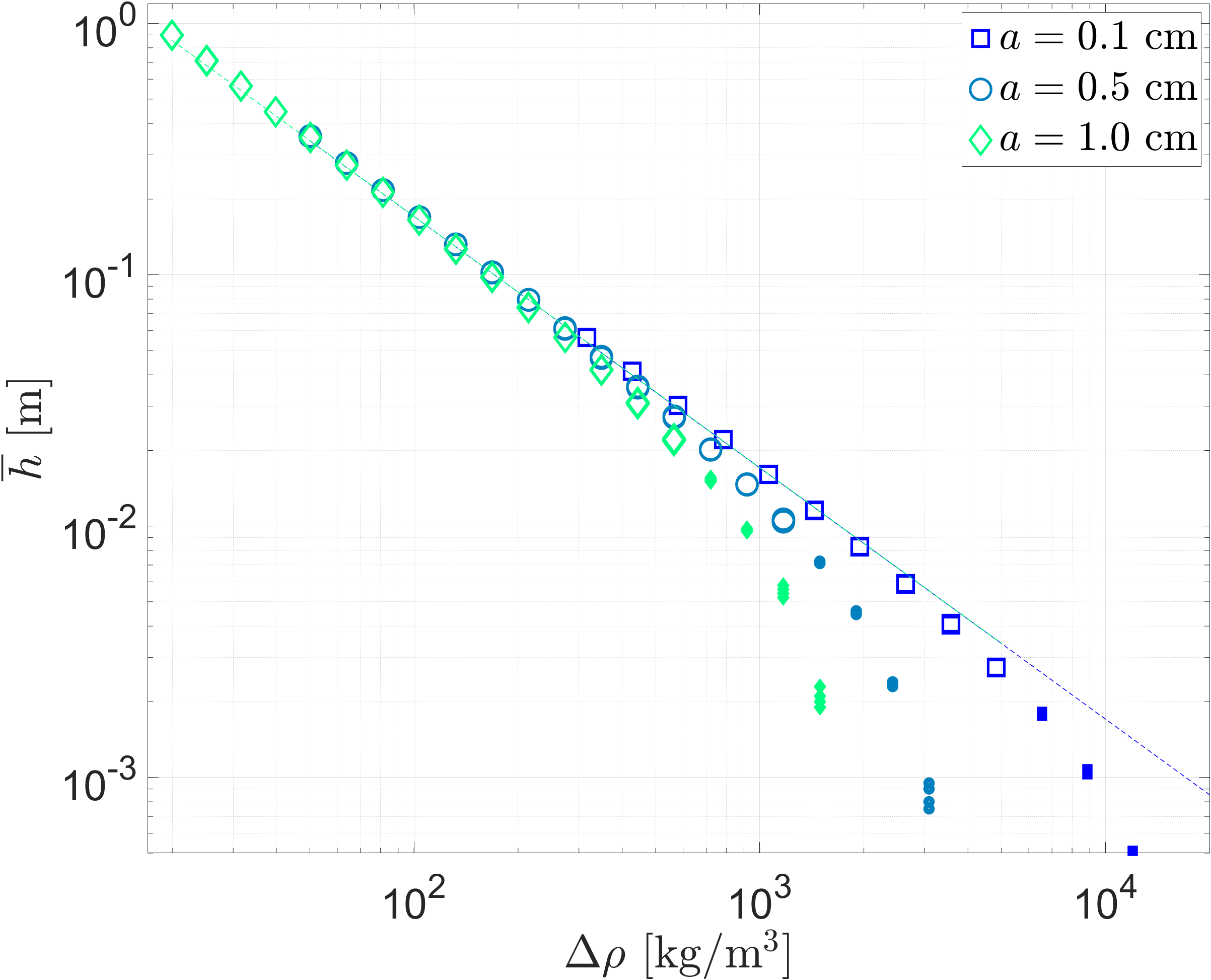}
         \caption{Diamond Lattice - Uniform $N(P_t)$}
         \label{fig:h_4_0}
     \end{subfigure}
     \hfill
     \begin{subfigure}[t]{0.48\textwidth}
         \centering
         \includegraphics[width=1\textwidth]{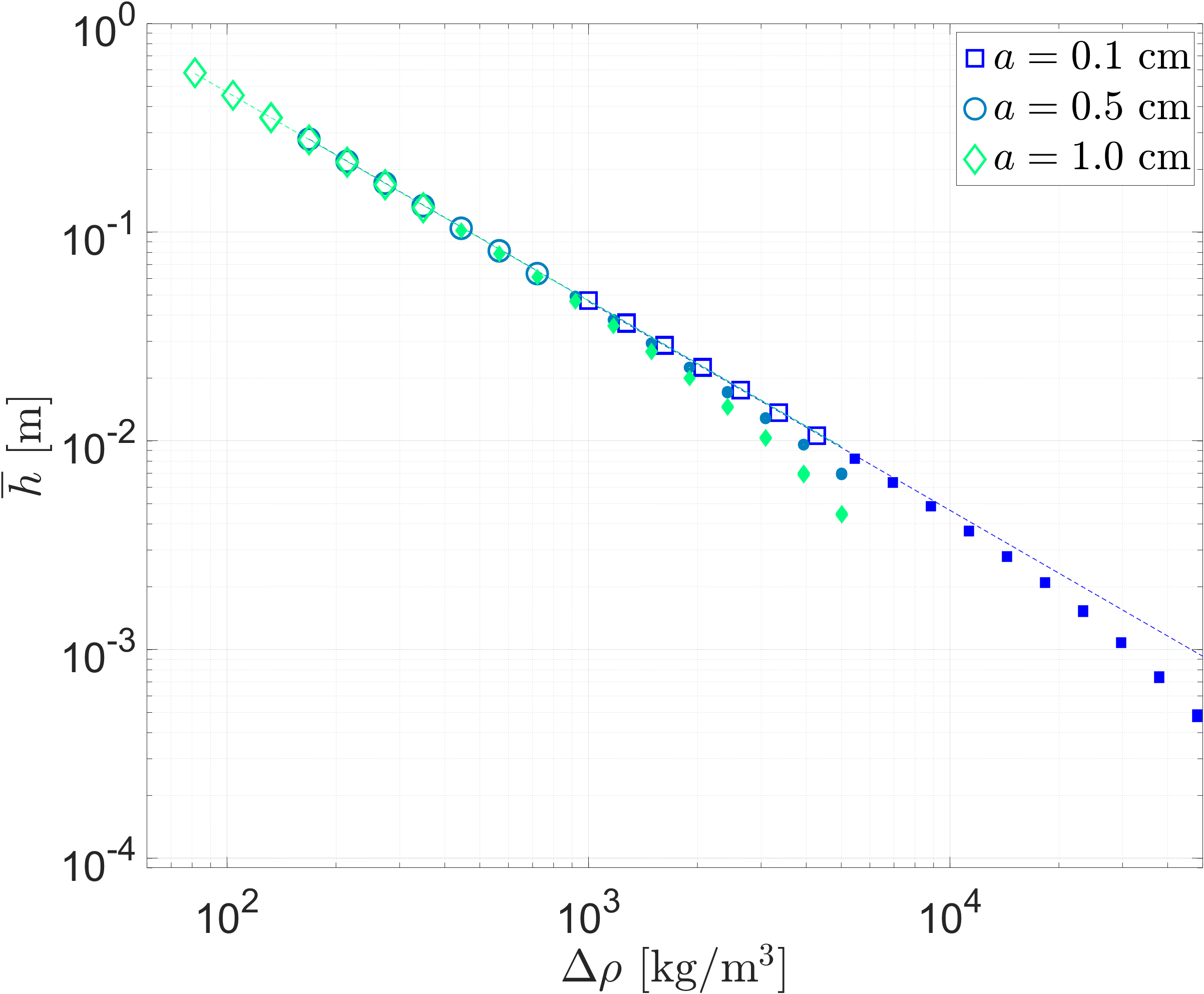}
         \caption{Simple Cubic Lattice - Non-uniform $N(P_t)$}
         \label{fig:h_1_2}
     \end{subfigure}
     \hfill
     \begin{subfigure}[t]{0.48\textwidth}
         \centering
         \includegraphics[width=1\textwidth]{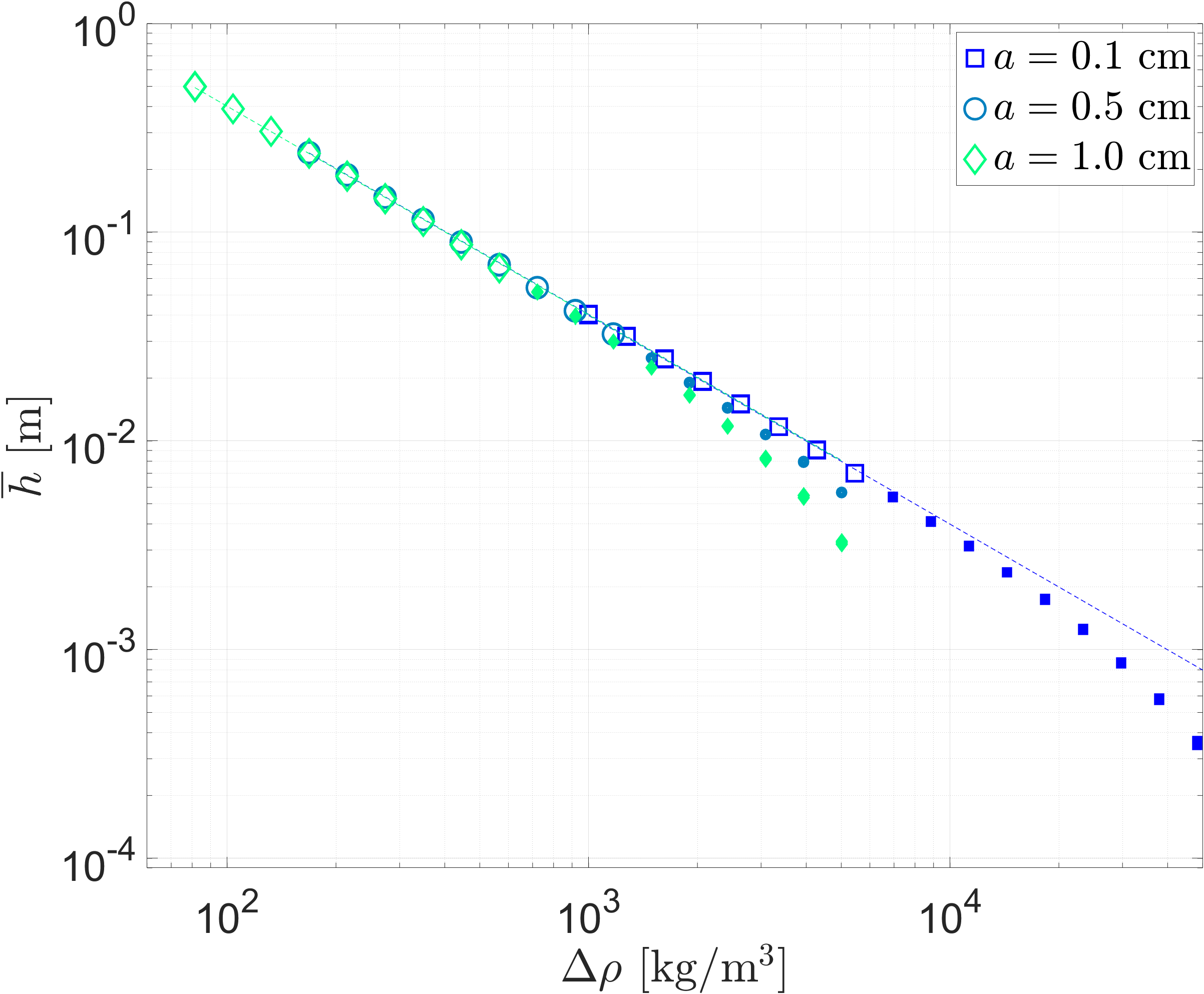}
         \caption{Simple Cubic Lattice - Uniform $N(P_t)$}
         \label{fig:h_1_0}
     \end{subfigure}
     \hfill

        \caption{Measured widths of the transition zone $h$ in IP drainage simulations, for all tested pore-networks. Big hollow symbols represent data within the range of simulation parameters we expect the theoretical framework presented in Sec. \ref{sec:theo} to be valid. Small filled symbols are shown otherwise. The dashed lines represent the expected values of $h=(P_{res}-P_{crit})|G|^{-1}$.}
        \label{fig:front_center}
\end{figure}

Figure \ref{fig:front_center} presents the measured values of $h$ using this procedure, for the two pore-network topologies, two distributions of capillary pressure threshold values, and three average pore sizes presented in Sec. \ref{sec:IP}. Results from five random realizations of each different pore-network type are shown. Values of $h$ are plotted against the difference in density between the phases, $\Delta \rho$, used in the drainage simulations. As discussed in Sec. \ref{sec:theo}, we expect $h$ to be equivalent to $(P_{crit}-P_{res})|G|^{-1}$. Therefore, in the case of gravity stabilized drainage, the transition width between the front's critical regions should be inversely proportional to $\Delta \rho$. 

From Fig. \ref{fig:h_4_2} to \ref{fig:h_1_0}, green diamond symbols represent $a=1.0$ cm, light blue circles represent $a=0.5$ cm, and dark blue squares represent $a=0.1$ cm. The results show that the transition zone widths obtained with the IP model match the expected values for all tested pore-network types, if $\Delta\rho$ values are not too large. In this regard, we observe that curves from pore networks with larger $a$ deviate from $h=(P_{crit}-P_{res})|G|^{-1}$ at lower $\Delta \rho$ values. This suggests that partitioning the front into a transition zone and critical regions may not be straightforward when the pressure gradients are too high and the front spans only a few pores.

For this reason, a brief investigation of the effect of high gradients on the proposed front scaling is presented in  Appendices \ref{app:nabla_p} and 
\ref{app:nabla_p_drainage}. Based on that, we verify that the theoretical framework presented in Sec. \ref{sec:theo} may not be valid if the density difference between the phases is larger than $\Delta \rho^{lim}=(|\nabla p|a)^{lim}/(N(P_{crit})ga)$, where $|\nabla p|a$ invasion probability difference over the length of a bond at the front, with a suggested limit of $(|\nabla p|a)^{lim}=0.075$. To make this point clear, results obtained with $\Delta \rho<\Delta\rho^{lim}$ are represented with big hollow symbols, while smaller filled symbols are used otherwise. Although this criterion is empirically based on gradient percolation results, it seems to predict the limit of $\Delta\rho$ where measured $h$ values fit the theoretical prediction adequately. Therefore, we keep this symbol distinction in the results presented in the following sections.

\subsection{Drainage front critical region near $P=P_{crit}$}
\label{sec:tail_crit}

In Figs. \ref{fig:nodes_center_example} and \ref{fig:pc_vs_z_invaded}, it is noticeable that a significant fraction of the drainage front is located outside the transition zone $h$. At the front tip, where the capillary pressure is near $P_{crit}$, the extent of the front below $h$ corresponds to the critical region $\eta_t$, defined in Sec. \ref{sec:theo}. To evaluate the scaling of the width of this region, values of $\eta_t$ are calculated as 

\begin{equation}
        \eta_t^{2}= \sum_{i=1}^{N_t} \frac{(z_i-z_{c})^2}{N_t} 
    \label{eq:Gouyet_eta_t}
\end{equation}

\noindent{ where $N_t$ is the total number of invaded bonds connected to the front at $p<p_c$}, $z_i$ is the position of these bonds, and $z_{c}$ is the position of the lower limit of the transition zone $h$, at which the maximum allowed $P_t$ for invasion is $P_{crit}$ and $p=p_c$ (see Fig. \ref{fig:pc_vs_z_invaded}).

\begin{figure}[ht!]
     \centering
     
     \begin{subfigure}[t]{0.48\textwidth}
         \centering
         \includegraphics[width=1\textwidth]{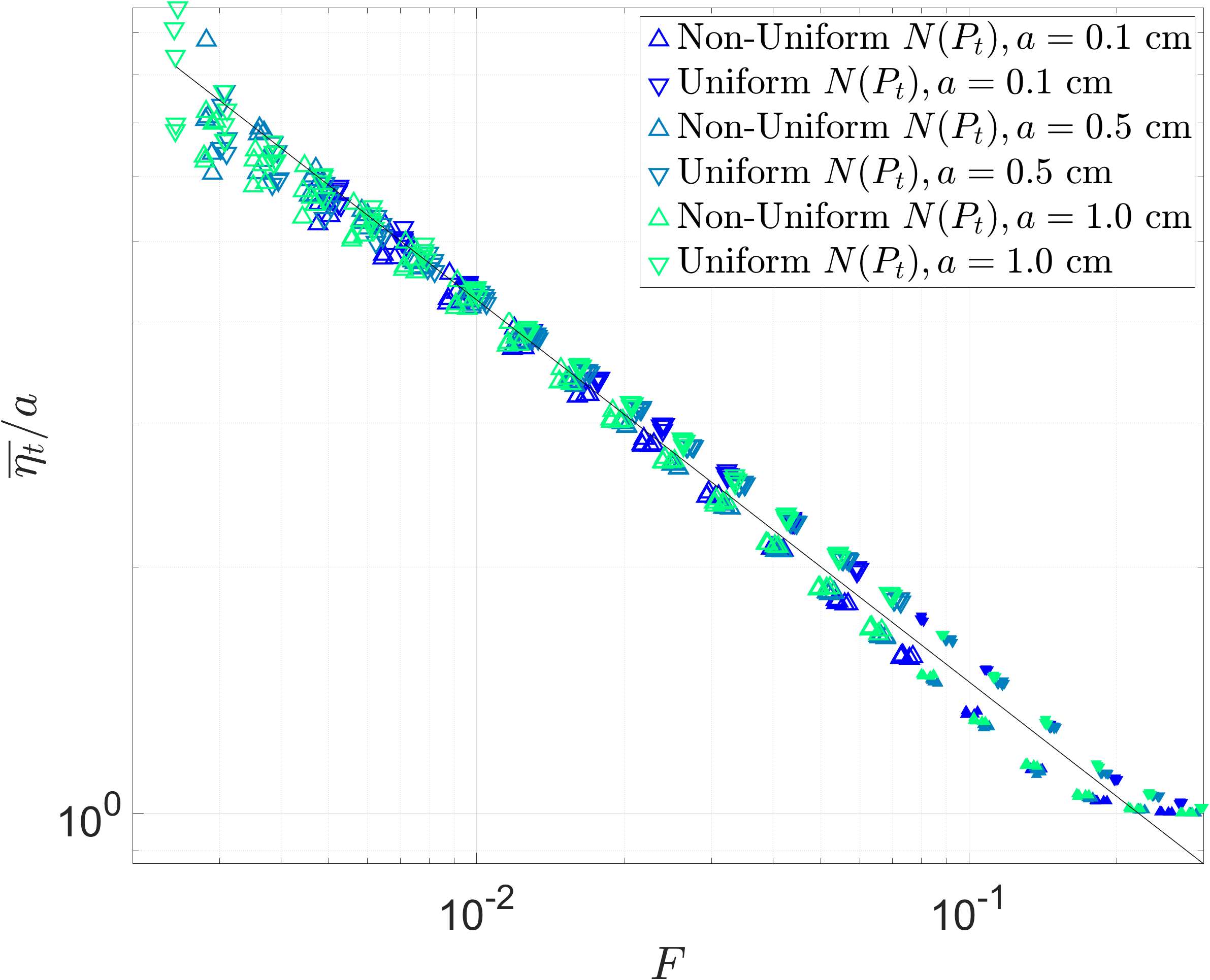}
         \caption{Diamond-Cubic Network}
         \label{fig:t_crit_4}
     \end{subfigure}     
     \hfill
     \begin{subfigure}[t]{0.48\textwidth}
         \centering
         \includegraphics[width=1\textwidth]{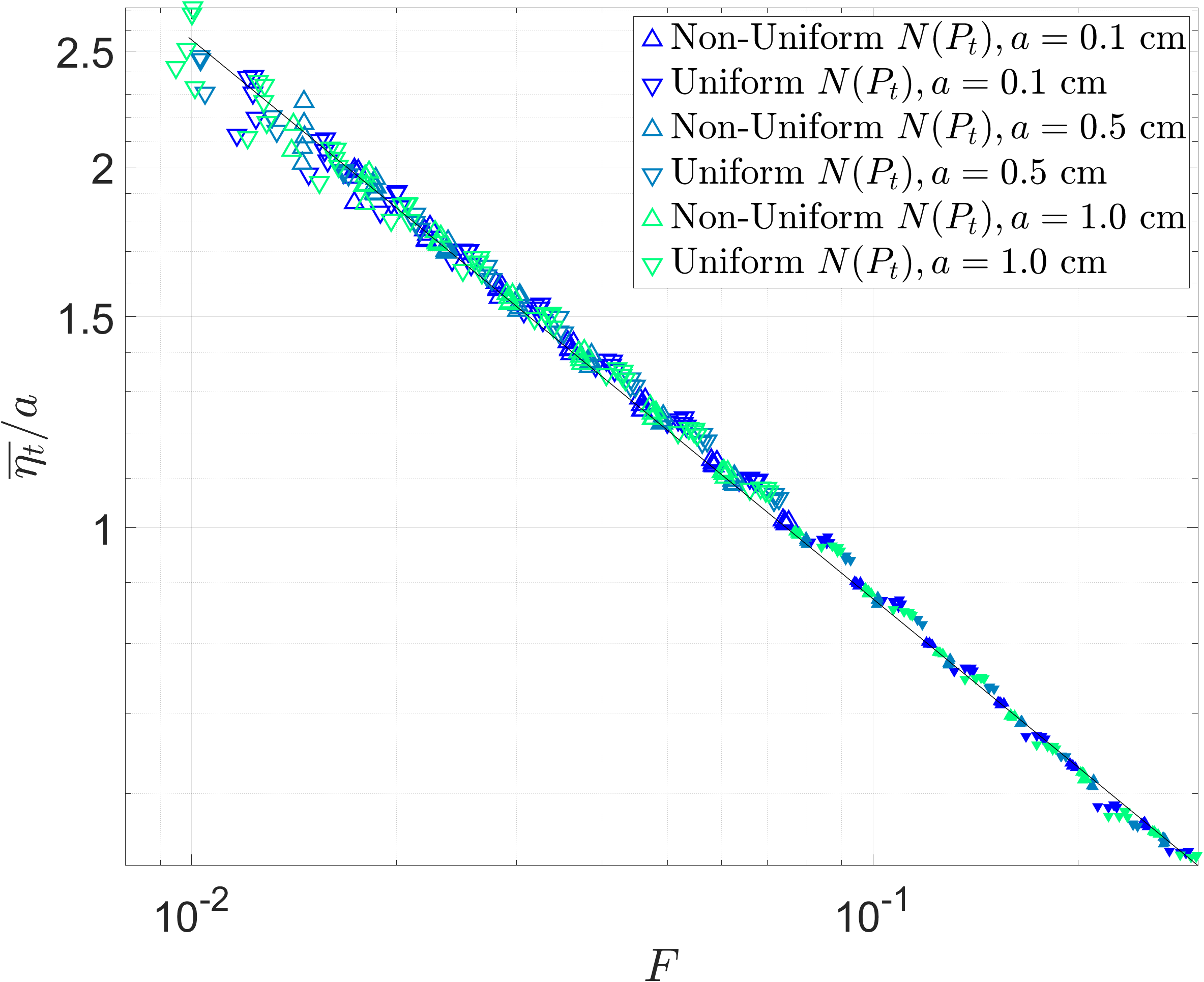}
         \caption{Simple-Cubic Network}
         \label{fig:t_crit_1}
     \end{subfigure}
     \hfill
        \caption{Calculated values of $\overline{\eta_t}/a$ in IP drainage simulations using Eq. \ref{eq:Gouyet_eta_t}, for all tested pore-networks. Big hollow symbols represent data within the range of simulation parameters we expect the theoretical framework presented in Sec. \ref{sec:theo} to be valid. Small filled symbols are shown otherwise. The black lines represent the expected scaling of $\eta_t/a\propto F^{-\nu/(1+\nu)}$, where $\nu=0.88$ from 3D percolation}.
        \label{fig:tail_crit}
\end{figure}

Instead of directly measuring the total extent of the front where $p<p_c$, Eq. \ref{eq:Gouyet_eta_t} provides a more statistically relevant quantification of how far the invaded bonds in the critical region spread from the position where $p=p_c$. Analogous to computing a standard deviation, this method was proposed by \citet{gouyet1988fractal} to verify the scaling of the infinite cluster front tail in gradient percolation. In Fig. \ref{fig:tail_crit}, values of $\overline{\eta_t}/a$ (where $\overline{\eta_t}$ is the time-averaged value of $\eta_t$) are plotted against $F=-N(P_{crit})Ga$. Results from five random realizations of each type of pore network are shown. 

The results from Fig. \ref{fig:tail_crit} clearly indicate that the values of $\eta_t/a$ obtained with the bond IP simulations agree well with the theoretical scaling based on gradient percolation, shown in the black lines. As expected, we see that values of $\eta_t$ depend on the average pore size $a$, unlike the transition zone $h$. Comparing the results obtained with the simple cubic and diamond-cubic pore networks, we can also see that values of $\eta_t$ differ even when $F$ and $a$ are the same. The observed differences are attributed to the scaling prefactor, which likely depends on the pore network topology. In Appendix \ref{app:const_c}, we propose a simple way to estimate the prefactor values, based on the front tail extent in gradient percolation. 

\subsection{Drainage front critical region near $P=P_{res}$}
\label{sec:tail_res}

Similarly to $\eta_t$, we propose that a second portion of the drainage front displays a critical behavior, as the occupation of the defending phase is near its percolation threshold. The extent of this region, termed $\eta_r$ in Sec \ref{sec:theo}, is illustrated in Fig. \ref{fig:nodes_center_example} above the upper limit of the transition zone $h$. Here, we investigate the scaling of $\eta_r$ with a procedure analogous to that presented in Sec. \ref{sec:tail_crit}.

\begin{figure}[ht!]
     \centering
     
     \begin{subfigure}[t]{0.48\textwidth}
         \centering
         \includegraphics[width=1\textwidth]{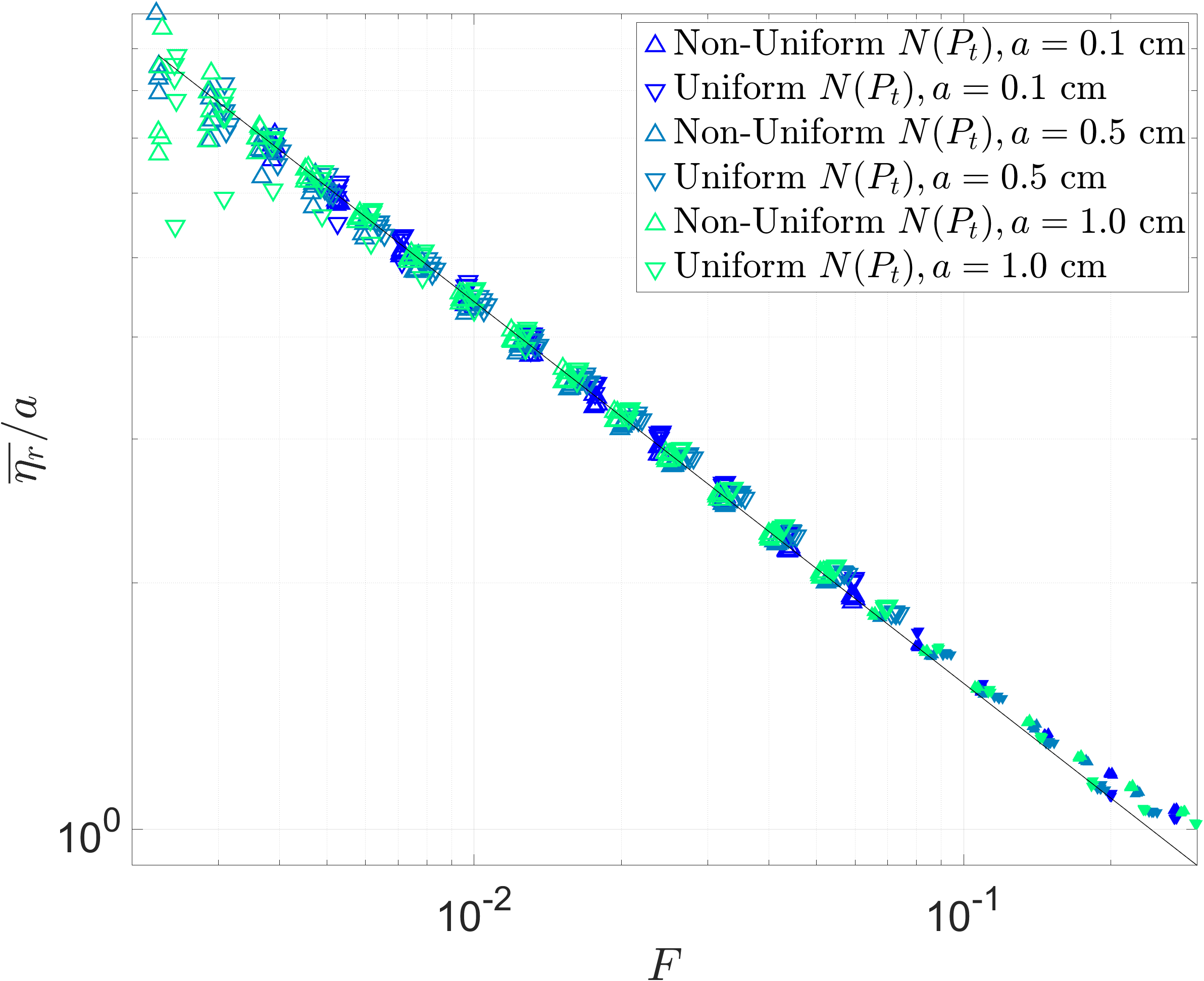}
         \caption{Diamond-Cubic Network}
         \label{fig:t_max_4}
     \end{subfigure}
     \hfill
     \begin{subfigure}[t]{0.48\textwidth}
         \centering
         \includegraphics[width=1\textwidth]{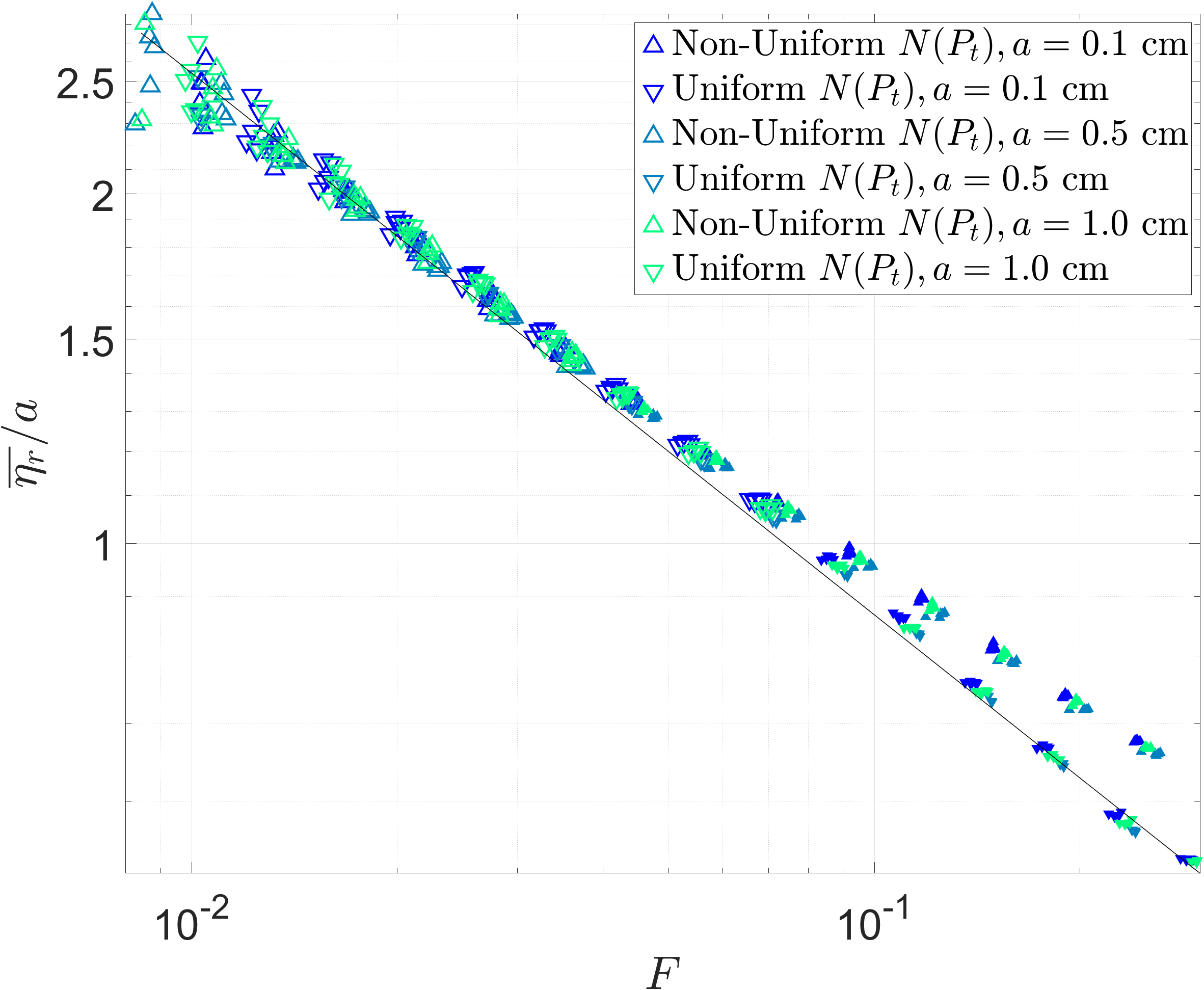}
         \caption{Simple-Cubic Network}
         \label{fig:t_max_1}
     \end{subfigure}
     \hfill

        \caption{Calculated values of $\overline{\eta_r}/a$ in IP drainage simulations using Eq. \ref{eq:Gouyet_eta_r}, for all tested pore-networks. Big hollow symbols represent data within the range of simulation parameters we expect the theoretical framework presented in Sec. \ref{sec:theo} to be valid. Small filled symbols are shown otherwise. The black lines represent the expected scaling of $\eta_r/a\propto F_r^{-\nu/(1+\nu)}$, where $\nu=0.88$ from 3D percolation}.
        \label{fig:tail_max}
\end{figure}

With Eq. \ref{eq:Gouyet_eta_r}, we calculate $\eta_r$ during drainage in five random realizations of each pore-network type presented in Sec. \ref{sec:nets}. In Fig. \ref{fig:tail_max}, values of $\overline{\eta_r}/a$ are plotted against $F_r=-N(P_{res})Ga$, defined as a modified fluctuation number. We observe that the extent of the drainage front where the defending phase becomes trapped in cluster scales as theoretically predicted, for all types of pore networks investigated. Along with the scaling of $\eta_t$ and $h$, these results strongly indicate that the total width of 3D stable drainage fronts may be correctly estimated with Eq. \ref{eq:FF_3D_front}.

\begin{equation}
        \eta_r^2= \sum_{i=1}^{N_r} \frac{(z_i-z_{r})^2}{N_r} 
    \label{eq:Gouyet_eta_r}
\end{equation}

\noindent{where $N_r$ is the total number of non-invaded bonds connected to the front at $p>1-p_c$}, $z_i$ is the position of these bonds, and $z_{r}$ is the position of the upper limit of the transition zone $h$, at which the maximum allowed $P_t$ for invasion is $P_{res}$ and $p=1-p_c$ (see Fig. \ref{fig:pc_vs_z_invaded}).

\subsection{Maximum length of trapped clusters}
\label{sec:L_max}

In two-dimensional gradient percolation, \citet{sapoval1985fractal} proposed that the width of the infinite cluster front should scale as the maximum length of clusters formed in its vicinity. The same relation was observed between the invasion front width and the maximum length of wetting-phase trapped clusters in gradient-stabilized drainage in 2D porous media \cite{birovljev1991gravity,khobaib2025gravity}. In three-dimensional gradient percolation, however, \citet{gouyet1988fractal} verified that the infinite-cluster front width is much larger than the clusters formed alongside it. Instead, they proposed that the length of these clusters is limited by the front tail width, with $\eta_{t}\propto|\nabla p|^{-\nu/(1+\nu)}$. Using a 3D invasion-percolation model with buoyancy effects, \citet{wilkinson1984percolation} similarly verified that the maximum length of trapped defending-phase clusters scaled as $L_{max}\propto Bo^{-\nu/(1+\nu)}$, where $Bo$ is the Bond number.

\begin{figure}[ht!]
     \centering
     \begin{subfigure}[t]{0.48\textwidth}
         \centering
         \includegraphics[width=1\textwidth]{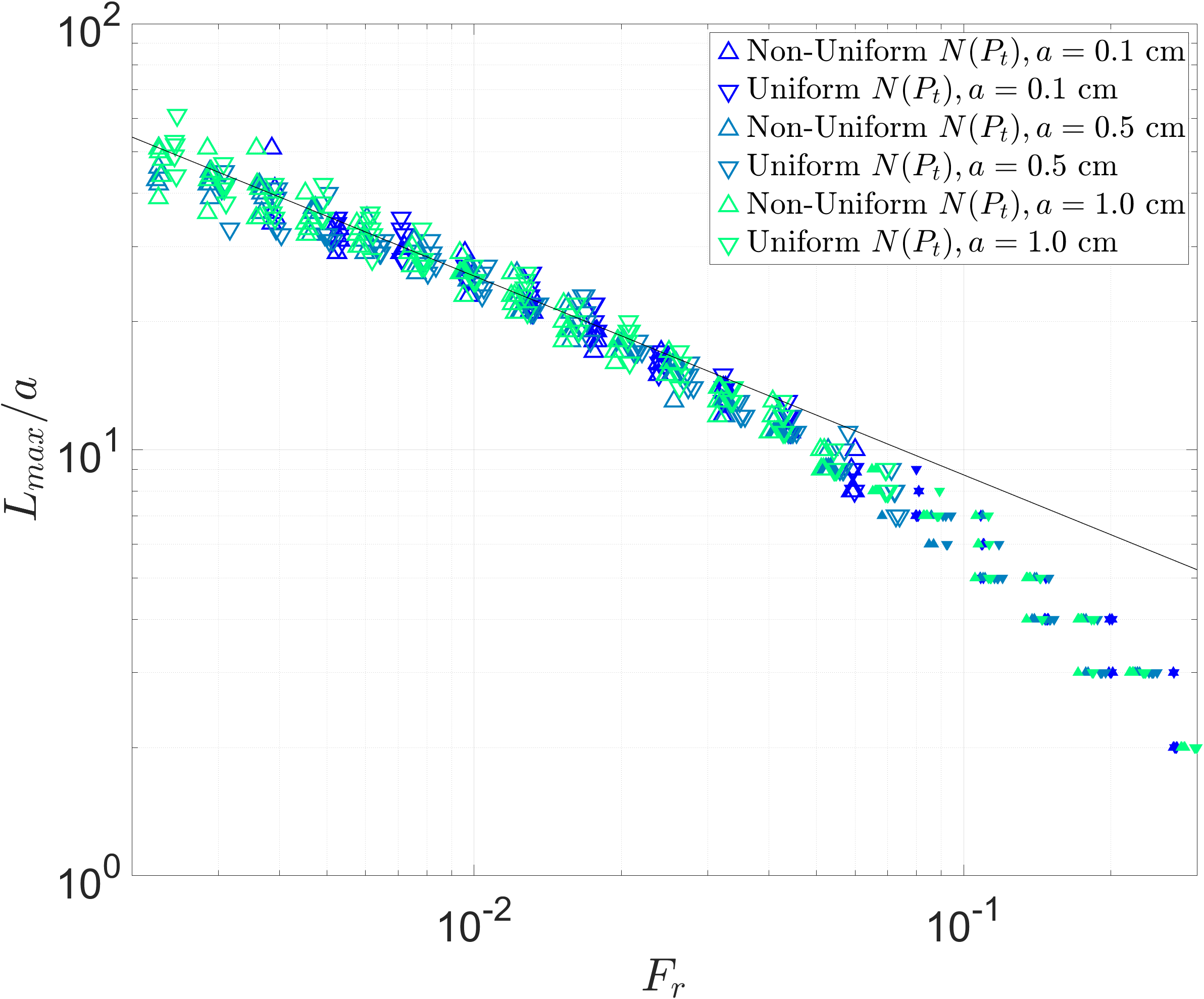}
         \caption{Diamond-Cubic Network}
         \label{fig:L_max_4}
     \end{subfigure}
     \hfill
     \begin{subfigure}[t]{0.48\textwidth}
         \centering
         \includegraphics[width=1\textwidth]{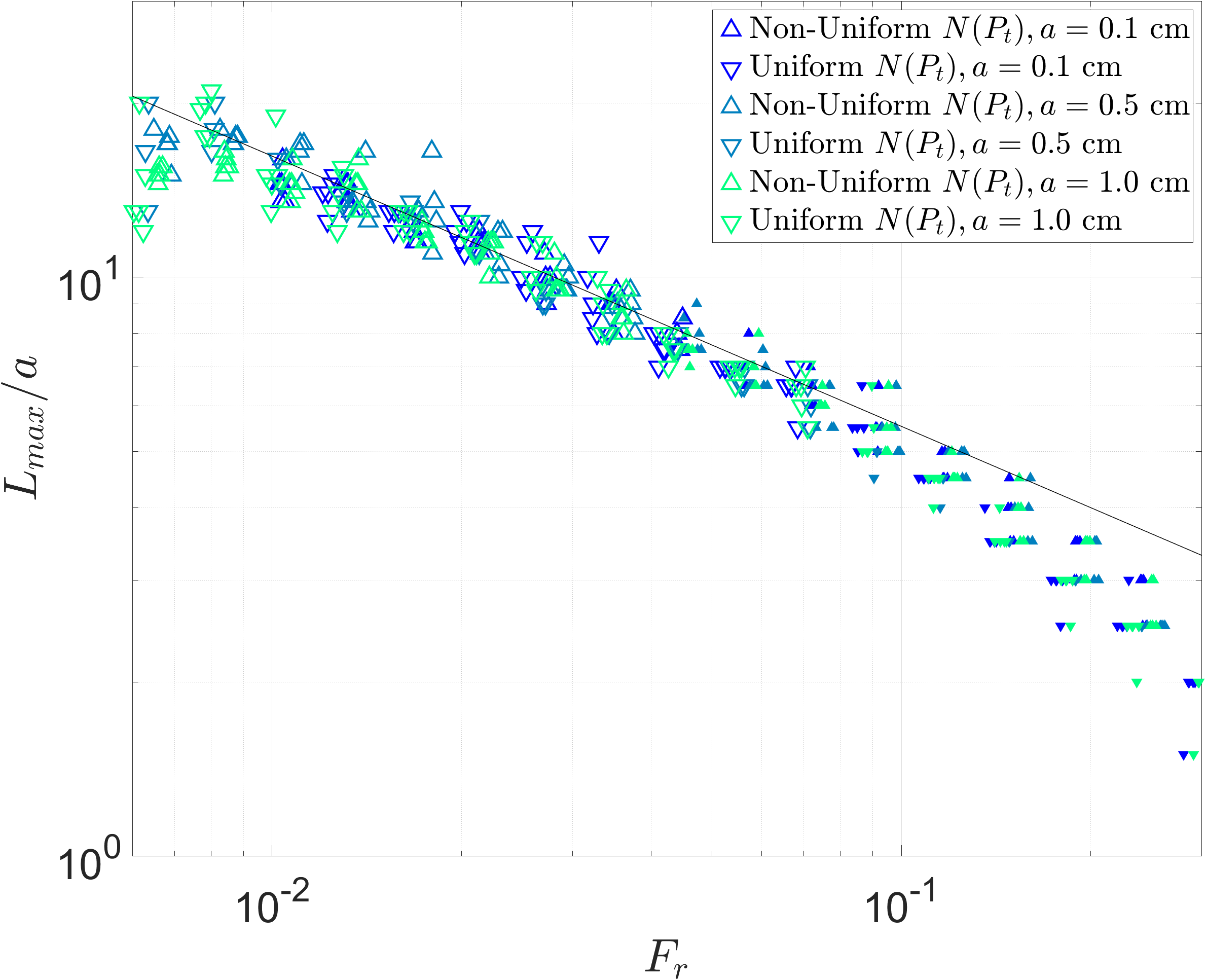}
         \caption{Simple-Cubic Network}
         \label{fig:L_max_1}
     \end{subfigure}
     \hfill

        \caption{Values of $L_{max}/a$ measured in IP drainage simulations at breakthrough, for all tested pore-networks. Big hollow symbols represent data within the range of simulation parameters we expect the theoretical framework presented in Sec. \ref{sec:theo} to be valid. Small filled symbols are shown otherwise. The black lines represent the expected scaling of $L_{max}/a\propto F_r^{-\nu/(1+\nu)}$, where $\nu=0.88$ from 3D percolation}.
        \label{fig:L_max}
\end{figure}

In Fig. \ref{fig:L_max}, we present the maximum length of trapped clusters normalized by the average pore size at breakthrough in our IP drainage simulations. Results are plotted against $F_r$, as these clusters are formed in the critical region where the defending phase is at its critical occupation. Therefore, we expect $L_{max}$ to scale as $\eta_r$.

Overall, the observed maximum cluster length scale reasonably well as $L_{max}/a\propto F_r^{-\nu/(1+\nu)}$. For high values of $F_r$, the tendency of $L_{max}$ to fall short of the theoretically predicted values may be attributed to the high gradient of occupation probability, as discussed in Appendices \ref{app:nabla_p} and \ref{app:nabla_p_drainage}. In fact, we notice that in these cases $L_{max}$ corresponds to the length of only a few pores, which can negatively impact the expected scaling. For very low $F_r$ values, we also identify that $L_{max}$ deviates from the prediction, especially when the simple-cubic pore network is used. In this scenario, the length of the drainage front (see Sec. \ref{sec:front_width}) is very close to the total length of the networks, and very few trapped clusters can be verified at breakthrough. Therefore, larger simulated domains may be required to estimate $L_{max}$ when $\eta_{3D}\approx an_z$.

\subsection{Drainage front width}
\label{sec:front_width}

In Sec. \ref{sec:theo}, we propose that pressure-stabilized drainage fronts in 3D random porous media can be partitioned into three sections: a critical region $\eta_t$ where the invading phase is near its percolation threshold, a second critical region $\eta_r$ where the defending phase is near percolation, and a transition zone $h$ where both phases can percolate. In Secs. \ref{sec:h}, \ref{sec:tail_crit}, and \ref{sec:tail_res}, we verify that these three regions satisfactorily scale according to the theoretical predictions based on gradient percolation. Here, we compare the total drainage front predicted with Eq. \ref{eq:FF_3D_front} with our bond IP model results.

\begin{figure}[ht!]
     \centering
     
     \begin{subfigure}[t]{0.48\textwidth}
         \centering
         \includegraphics[width=1\textwidth]{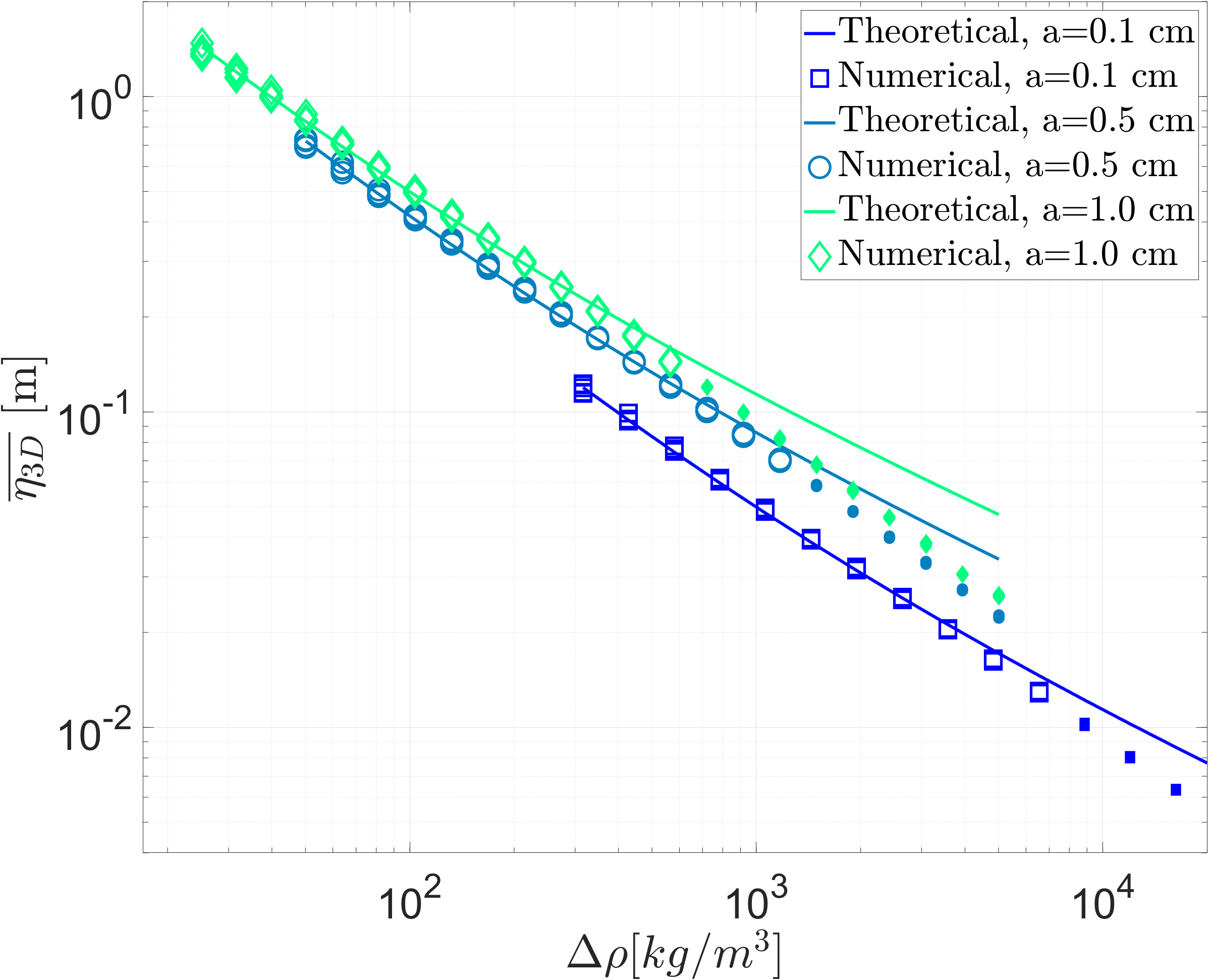}
         \caption{Diamond Lattice - Non-uniform $N(P_t)$}
         \label{fig:w_4_2}
     \end{subfigure}
     \hfill
     \begin{subfigure}[t]{0.48\textwidth}
         \centering
         \includegraphics[width=1\textwidth]{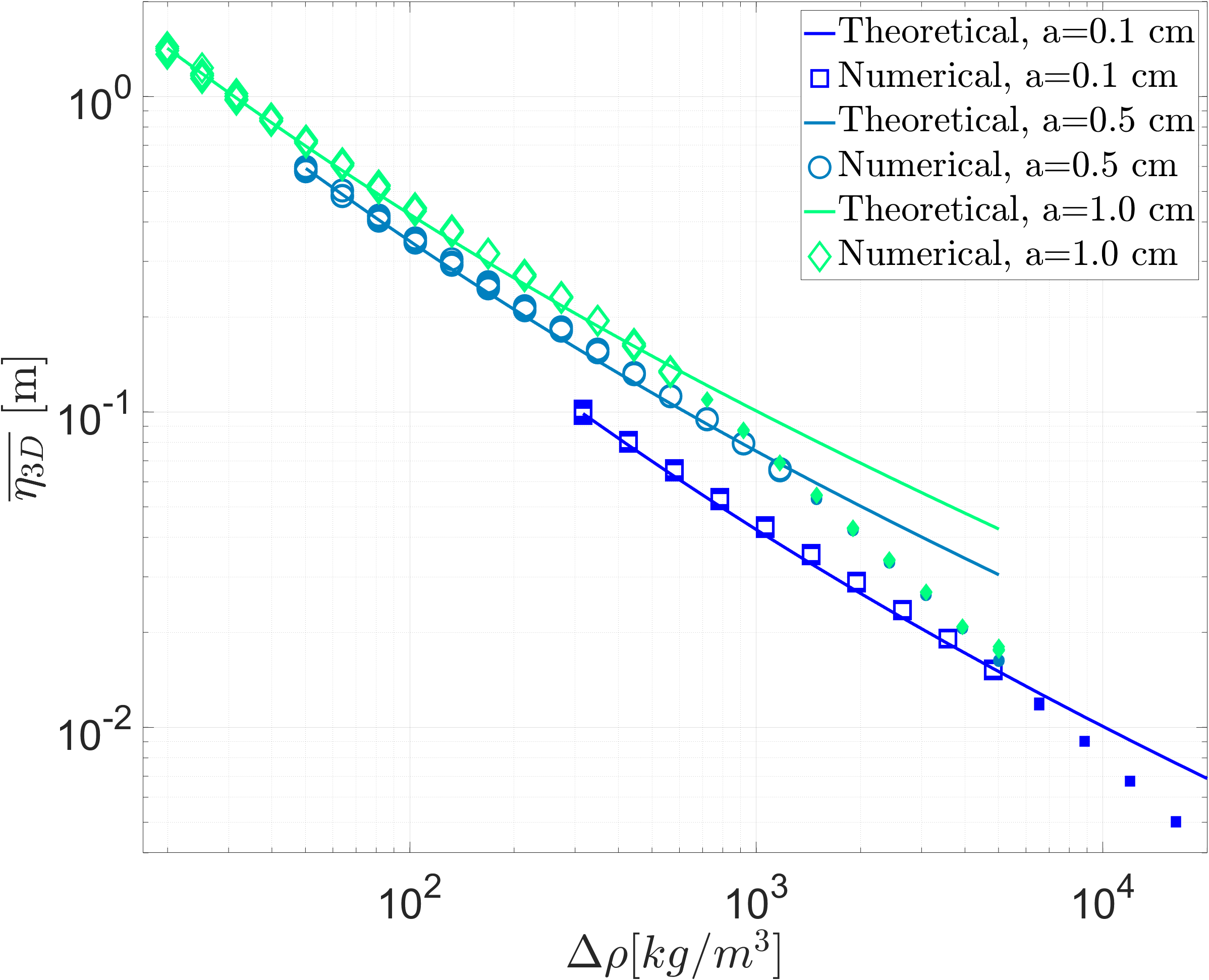}
         \caption{Diamond Lattice - Uniform $N(P_t)$}
         \label{fig:w_4_0}
     \end{subfigure}
     \hfill
     \begin{subfigure}[t]{0.48\textwidth}
         \centering
         \includegraphics[width=1\textwidth]{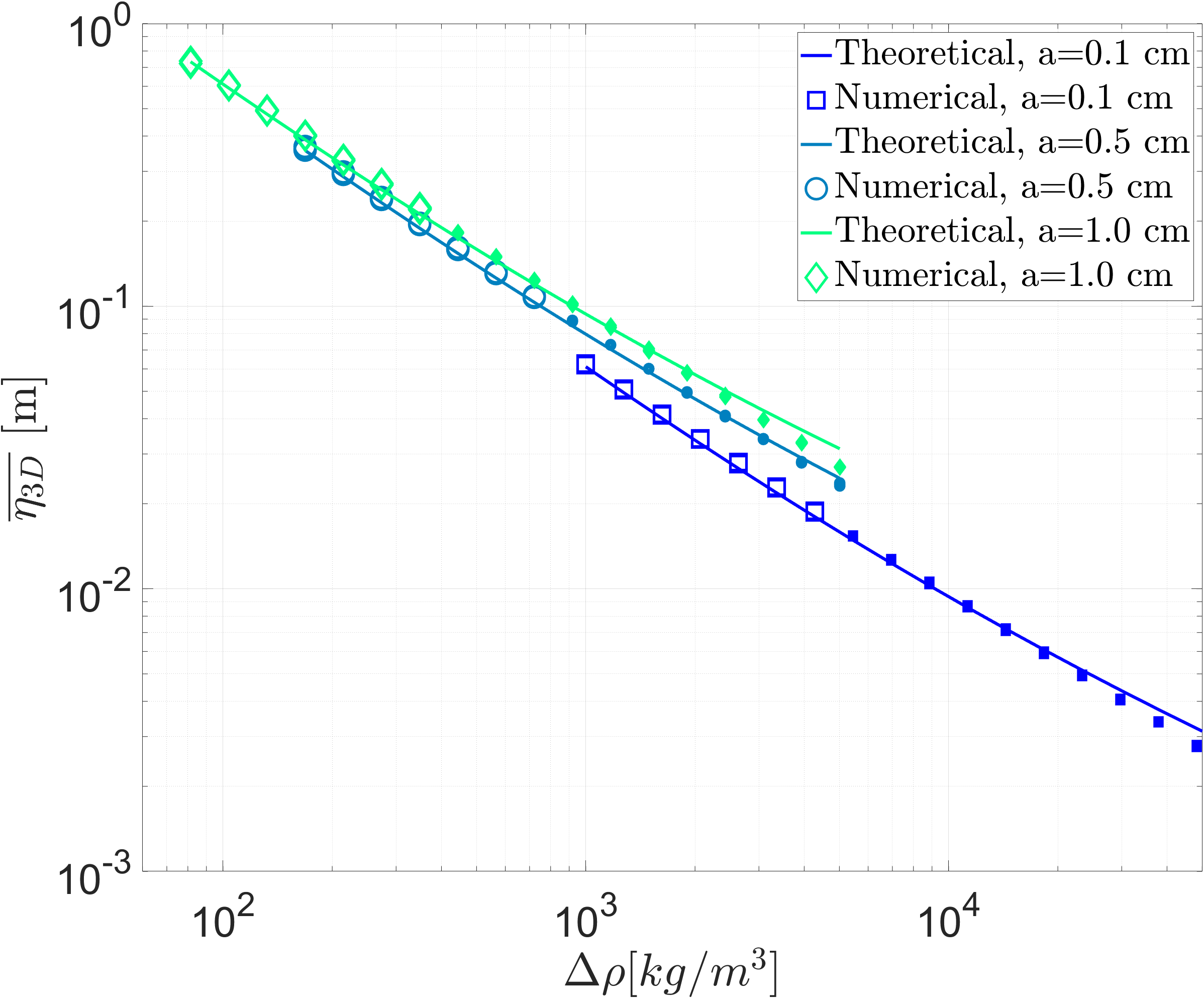}
         \caption{Simple Cubic Lattice - Non-uniform $N(P_t)$}
         \label{fig:w_1_2}
     \end{subfigure}
     \hfill
     \begin{subfigure}[t]{0.48\textwidth}
         \centering
         \includegraphics[width=1\textwidth]{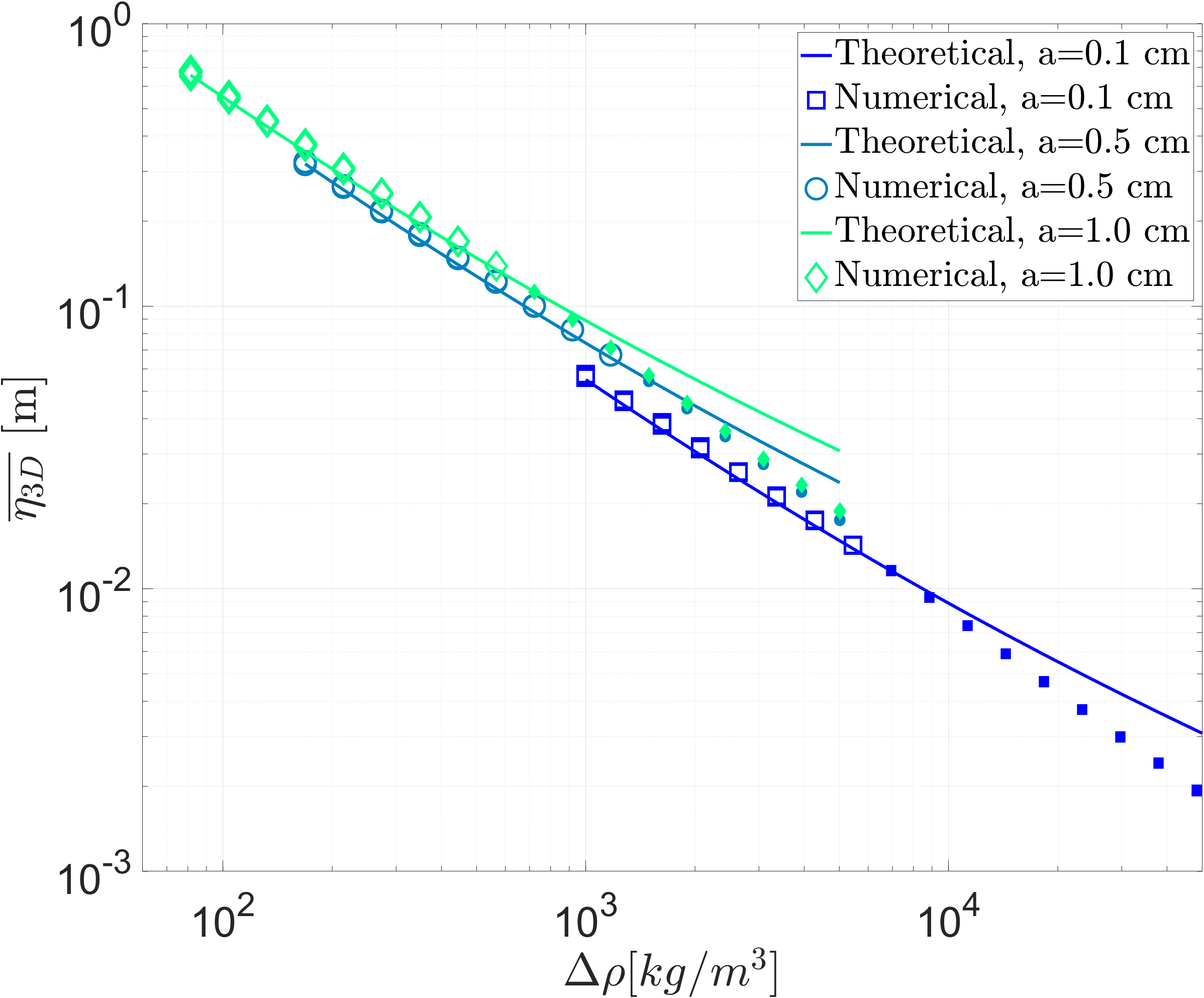}
         \caption{Simple Cubic Lattice - Uniform $N(P_t)$}
         \label{fig:w_1_0}
     \end{subfigure}
     \hfill

        \caption{Measured total front widths} $\eta_{3D}$ in IP drainage simulations, for all tested pore-networks. Big hollow symbols represent data within the range of simulation parameters we expect the theoretical framework presented in Sec. \ref{sec:theo} to be valid. Small filled symbols are shown otherwise. The continuous lines represent the estimated values of $\eta_{3D}$, using Eq. \ref{eq:FF_3D_front}.
        \label{fig:front_width}
\end{figure}

In Fig. \ref{fig:front_width}, the total front width measured with the drainage simulations is shown as scattered data, while the predictions from Eq. \ref{eq:FF_3D_front} correspond to the continuous lines. To use this equation, we adopted $\mathcal{C}=1.55$ for diamond-cubic pore networks and $\mathcal{C}=0.9$ for simple-cubic pore networks. These values are obtained in Appendix \ref{app:const_c}, based on gradient percolation, and seem to depend on the topology of the network only. The total front width is obtained from the simulations as simply $\eta_{3D}=z_\eta^{max}-z_\eta^{min}$, where $z_\eta^{max}$ and $z_\eta^{min}$ are the maximum and minimum values of $z$ among the front bonds, respectively.

We observe that the proposed estimate for $\eta_{3D}$ adequately fits the data obtained with numerical simulations, when values of $\Delta \rho$ are not too high. As presented in Appendices \ref{app:nabla_p} and \ref{app:nabla_p_drainage}, we expect this deviation to arise when the gradient of invasion probability along the front is large and the front spans only a few pores. In this scenario, it is verified in Sec. \ref{sec:h} that the prediction of the transition zone width $h=(P_{res}-P_{crit})|G|^{-1}$ also fails.

Conversely, when the total front width is larger than $\approx10a$, we demonstrate that Eq. \ref{eq:FF_3D_front} provides a good estimate of the stabilized drainage front widths obtained with a bond IP model. The correspondence between numerically and theoretically obtained $\eta_{3D}$ seems independent of the capillary pressure threshold distribution, average pore size, and network topology. It is noteworthy that no fit parameter is required for the front prediction.

\section{Discussion}
\label{sec:disc}

The results obtained with our bond invasion-percolation model with trapping suggest that gravity-stabilized drainage fronts in random porous media can be mapped onto the infinite-cluster frontier in gradient percolation. While both types of percolation do not belong to the same universality class \cite{wilkinson1984monte,vincent2022stable}, the capillary pressure gradient along the direction of the flow leads to a gradient in invasion probability along the drainage front analogous to the gradient in occupation probability in gradient percolation.

The partitioning of the front into two regions with critical behavior, $\eta_t$ and $\eta_r$, separated by a transition zone, $h$, is supported by individually assessing the scaling of each segment. Additionally, $\eta_{3D}$ predictions obtained with Eq. \ref{eq:FF_3D_front} are successfully compared with numerical results, when appropriate values of $G$ and $\mathcal{C}$ are used. Based on that, we demonstrate how stable drainage front widths in 3D porous media may have a more complex dependency on the stabilizing pressure gradients than in two dimensions.

Consider a random 3D porous medium undergoing stable drainage. At low $|G|$, the width of the front tends to be dominated by the width of the transition zone, which follows the scaling $h\propto|G|^{-1}$. At high $|G|$, $\eta_{3D}$ may become dominated by the width of the critical regions, which scale as $|G|^{-\nu/(1+\nu)}$, where ${-\nu/(1+\nu)}=-0.47$. Therefore, an attempt to simply establish a power-law relationship between $\eta_{3D}$ and $|G|$ may lead to fitted exponents between $-1$ and $-0.47$, depending on the evaluated range of $|G|$. Using a bond invasion-percolation model to represent gravity-stabilized drainage in a simple-cubic pore network, \citet{breen2022stable} reported a power-law fit equivalent to $\eta_{3D}\propto|G|^{-0.85}$, which is compatible with our observations.

The relative contribution of each segment to the front width also depends non-trivially on the pore structure. Highly connected networks exhibit low bond-percolation thresholds, leading to wider transition zones where both phases can percolate. In Fig. \ref{fig:front_center}, we notice that values of $h$ are larger in simple-cubic networks than diamond-cubic networks with the same $N(P_t)$ and $\Delta \rho$. Besides the pore-network topology, the width and form of the capillary pressure threshold distribution have a complex influence on $\eta_{3D}$. For a given pore network, values of $\eta_t$ and $\eta_r$ are equivalent if $N(P_t)$ is uniform. Using the non-uniform $N(P_t)$ illustrated in Fig. \ref{fig:MouraNPt}, $\eta_r$ is larger than $\eta_t$, as shown in Figs. \ref{fig:tail_crit} and \ref{fig:tail_max}, especially with the simple-cubic network.

Understanding how the fluids and porous medium properties affect the invasion front during drainage can be relevant for estimating important parameters in displacement flow studies, such as the representative elementary volumes (REV), the interfacial area, and the resulting residual saturation \cite{culligan2004interfacial}. Similar arguments to those presented in this work could also be useful in studies of drying fronts in porous media, whose shape and extent can control evaporation rates \cite{chen2017control}. 

Furthermore, the presented drainage front width analyses may be valid for a wide range of naturally occurring and engineered porous media, despite the suggested limitation in pressure gradient, when the characteristic pore size is in the order of micrometers. For example, consider the stable displacement of water by air in a Berea sandstone with properties presented in \citet{oren2003reconstruction}. With pore-throat radii, $r$, varying from 5 to 80 $\mu$m, and an average pore length, $a$, of approximately 200 $\mu$m, a ballpark estimate of the variation in invasion probability along one pore at the front is $|\nabla p|a=N(P_{crit})\Delta \rho ga\approx7.5\times10^{-5}$ (considering $g=10$ m/s$^2$, $\Delta \rho=1000$ kg/m$^3$, a uniform $N(P_t)$ and $P_t=2\gamma/r$, where $\gamma=72$ mN/m). This value is three orders of magnitude smaller than the limit of 0.075 identified in our simulations. 

Still, it is crucial that the theory presented in Sec. \ref{sec:theo} is validated experimentally. While IP models have been successfully used to predict drainage in two-dimensional porous media \cite{birovljev1991gravity,frette1992buoyancy,meakin2000invasion,vincent2022stable,reis2023simplified,khobaib2025gravity}, their ability to represent flow in three-dimensional porous media may be more limited. Also, assuming that no spatial correlations exist among capillary pressure thresholds in real 3D porous media may not be entirely realistic \cite{bryant1993network}. If that is the case, corrections for correlated disorder should be incorporated in the presented theoretical framework \cite{meakin1991invasion, mani1999effect}. Another interesting limitation to our analyses may arise in poorly connected 3D porous media. As reported by \citet{tran2013percolation}, 3D lattices with three nearest neighbors exhibit bond percolation thresholds $p_c\approx0.55$. Then, porous media with topology equivalent to these lattices could be percolated by only one phase at a time, similar to two-dimensional porous media. As a result, drainage front widths in such porous media may follow the scaling presented in Eq. \ref{eq:f_scaling_2D}. 

\section{Conclusion}
\label{sec:conc}

In this work, we presented a theoretical approach to estimate stable drainage front widths in three-dimensional random porous media under capillary and gravitational effects. Based on the infinite-cluster frontier in gradient percolation, we suggest that the extent of the interface between wetting and non-wetting phases during drainage exhibits a significantly more complex dependency on the porous medium structure and stabilizing pressure gradients in 3D than in 2D. This difference in behavior stems from the fact that there is a range of saturations in which two phases can percolate three-dimensional porous media together. In two dimensions, the wetting phase is trapped in clusters as the non-wetting phase reaches its percolation threshold, leading to narrower stable drainage fronts. 

The theoretical prediction of the front widths proposed in Eq. \ref{eq:FF_3D_front} was tested with a bond invasion-percolation model, which incorporates gravitational effects through a linear gradient of capillary pressure acting at the front. A satisfactory match between theoretical and numerical results was achieved when the gradient of invasion probability along the pore throats at the front was not too high. As a following step to these analyses, it is important that our results are experimentally verified.

\begin{acknowledgments}

We acknowledge the support of the University of Oslo, the Njord Centre, and the Research Council of Norway through the PoreLab Center of Excellence (project number 262644). We also thank Reanaud Toussaint, Marcel Moura, Gaute Linga, Per Arne Rikvold, and Eirik Grudde Flekk{\o}y for fruitful discussions.

\end{acknowledgments}

\appendix

\section{Effect of the occupation gradient magnitude in gradient percolation}
\label{app:nabla_p}

 The theoretical estimate of stable drainage front widths in Eq. \ref{eq:FF_3D_front} is based on characteristics of the infinite cluster in gradient percolation in 3D \cite{rosso1986gradient,gouyet1988fractal}. 
As proposed by \citet{gouyet1988fractal}, the extent of the infinite cluster front at $p<p_c$ is termed the front tail, and should scale with the occupation gradient $\nabla p$ as $\eta_{t}\propto|\nabla p|^{-\nu/(1+\nu)}$, where $\nu=0.88$ is the percolation critical exponent for the correlation length in 3D. In this Appendix, we analyze a series of gradient percolation cases to verify if the magnitude of $\nabla p$ affects this scaling. 

Similarly to our drainage simulations, diamond-cubic and simple-cubic networks are used. To match the size of the networks used in our IP model, $n_x=n_y=100$ for the diamond-cubic networks, while $n_x=n_y=50$ for simple-cubic networks. $n_z$ varied from 6 to 250 to obtain a large range of $\nabla p=1/n_z$ values. 

\begin{figure}[ht!]
     \centering
     
     \begin{subfigure}[t]{0.48\textwidth}
         \centering
         \includegraphics[width=1\textwidth]{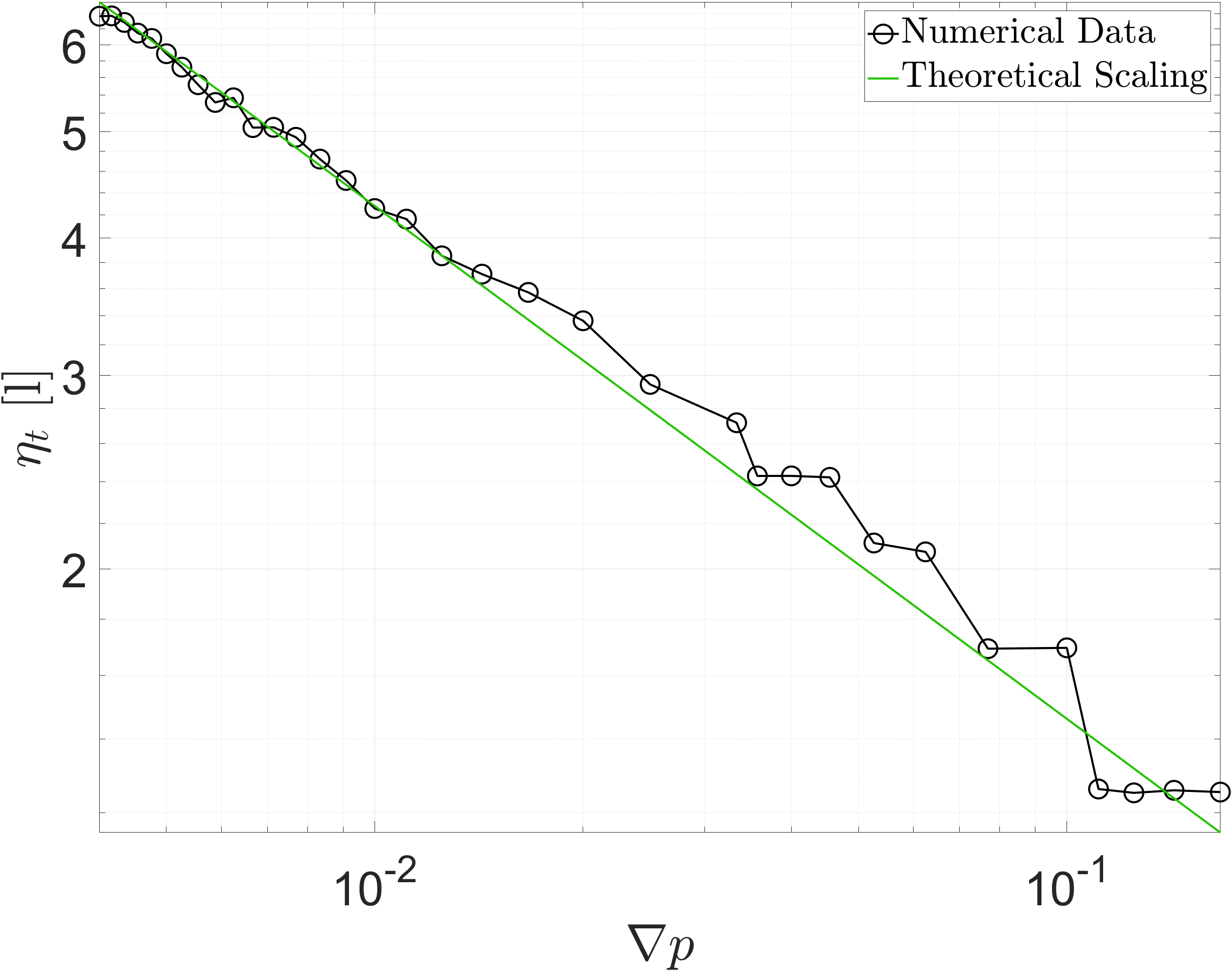}
         \caption{diamond-cubic network}
         \label{fig:eta_ft_Diamond}
     \end{subfigure}
     \begin{subfigure}[t]{0.48\textwidth}
         \centering
         \includegraphics[width=1\textwidth]{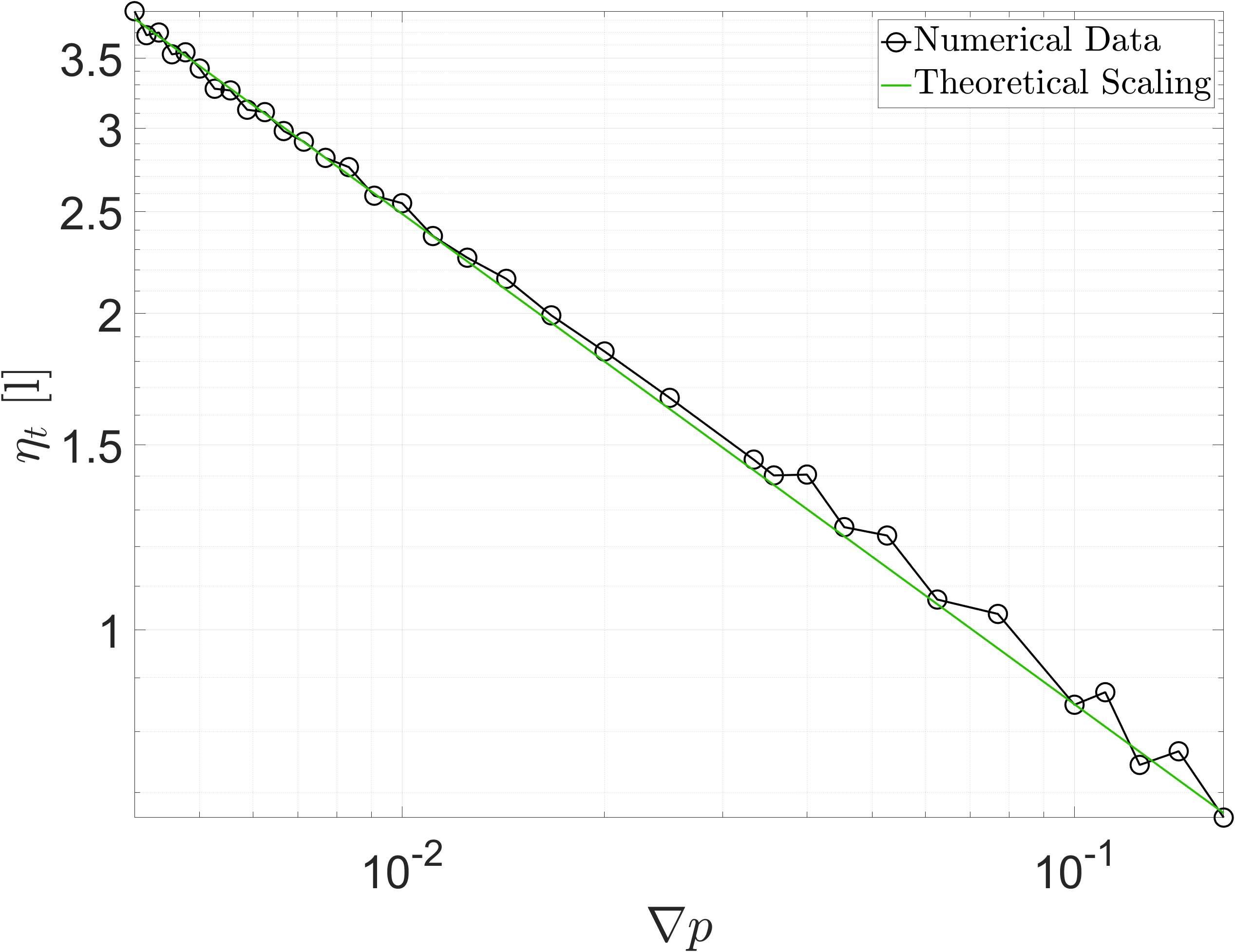}
         \caption{simple-cubic network}
         \label{fig:eta_ft_SC}
     \end{subfigure}

        \caption{Front tails width scaling in gradient percolation. The presented numerical results are calculated with Eq. \ref{eq:Gouyet_eta_t}, and are the average values from 100 realizations of random networks. The theoretical scaling curve shows the expected coefficient of $\beta=-\nu/(1+\nu)$, approximately equal to $-0.47$ in 3D.  Widths are presented in bond units $l$.}
        \label{fig:appendix_eta_ft}
\end{figure}

In Fig. \ref{fig:appendix_eta_ft}, we present the values of $\eta_{t}$ calculated with Eq. \ref{eq:Gouyet_eta_t}. The numerical results are the average from 100 random realizations of each type of network and value of $\nabla p$. A theoretical scaling curve is also shown in each plot, corresponding to the power-law coefficient of $\beta=-\nu/(1+\nu)$. We can see that the numerically obtained values of $\eta_{t}$ follow reasonably well the theoretical scaling in the range of $\nabla p$ evaluated. While higher values of $\nabla p$ impair the scaling of the front tails measured in the diamond-cubic networks, the calculated $\eta_{t}$ values still oscillate near the predicted values.

\begin{figure}[ht!]
     \centering
     
     \begin{subfigure}[t]{0.48\textwidth}
         \centering
         \includegraphics[width=1\textwidth]{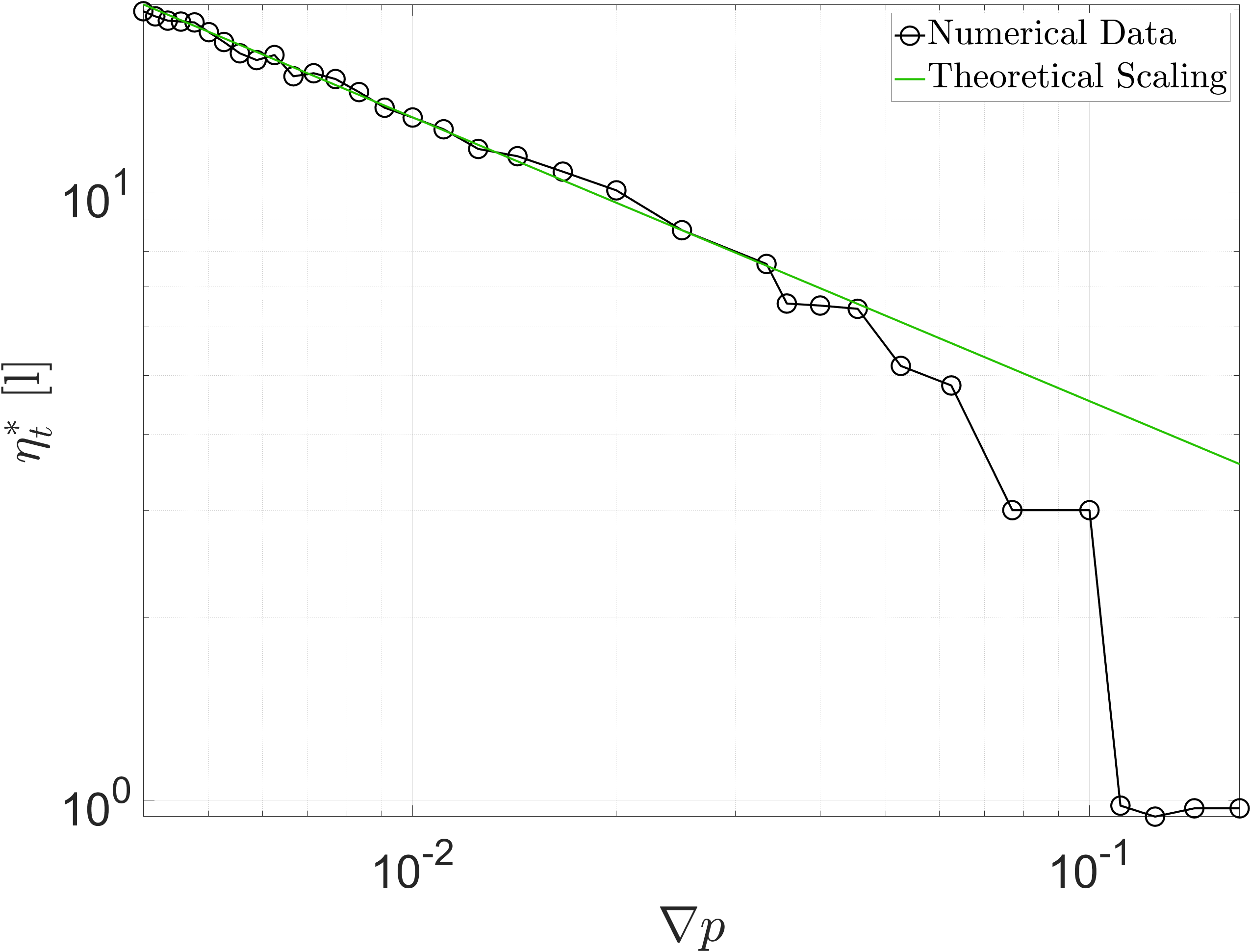}
         \caption{diamond-cubic network}
         \label{fig:eta_f_p_pc_Diamond}
     \end{subfigure}
     \begin{subfigure}[t]{0.48\textwidth}
         \centering
         \includegraphics[width=1\textwidth]{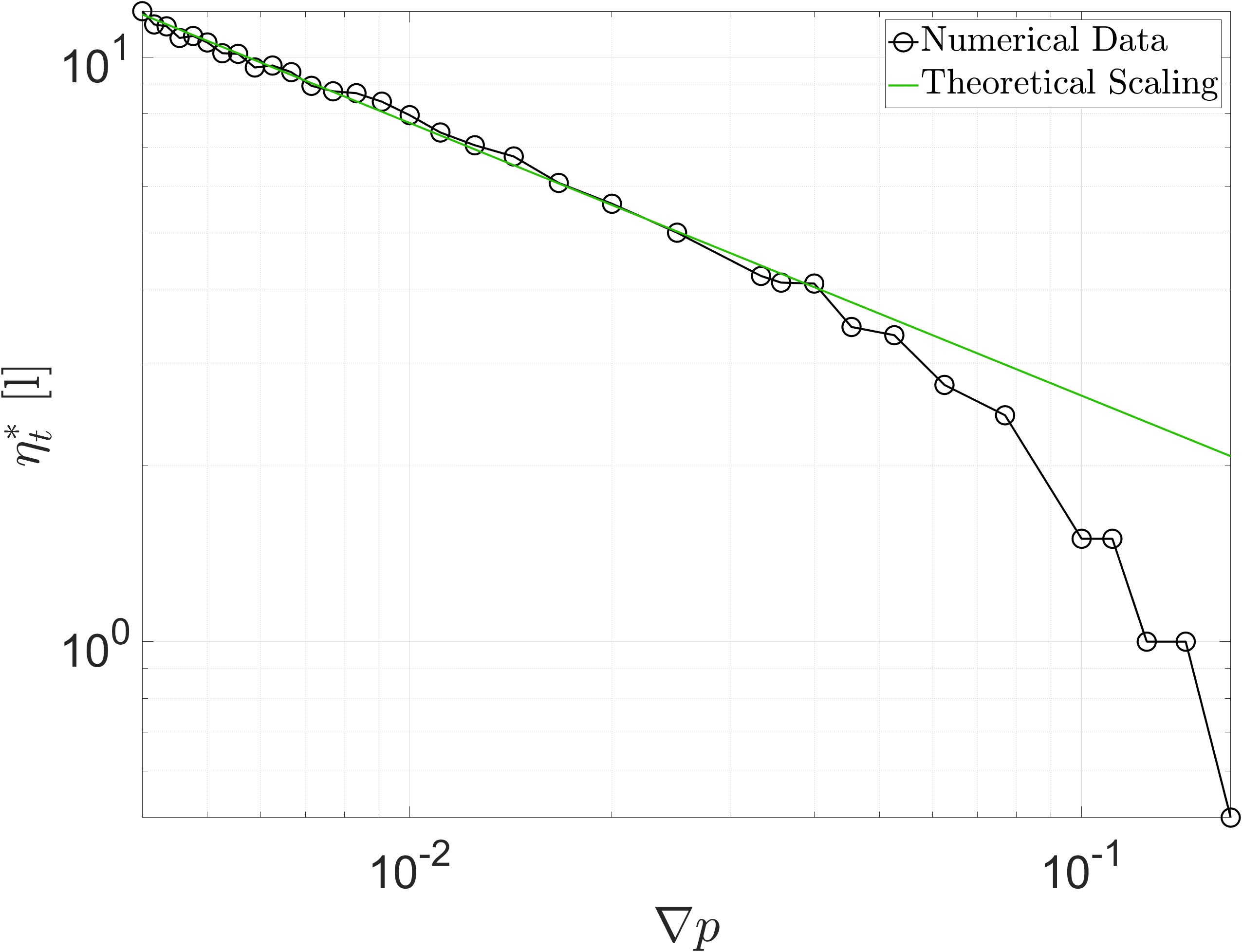}
         \caption{simple-cubic network}
         \label{fig:eta_f_p_pc_SC}
     \end{subfigure}

        \caption{Front tails width in gradient percolation, directly measured as the portion of the front $\eta_f$ where $p<p_c$: average values from 100 realizations of random networks. Widths are presented in bond units $l$.}
        \label{fig:appendix_eta_f_p_pc}
\end{figure}

In Fig. \ref{fig:appendix_eta_f_p_pc} we present results from the same gradient percolation cases as in Fig. \ref{fig:appendix_eta_ft}. However, now the front tail is measured directly as the total extent of the infinite cluster front in the region where $p<p_c$. Since in Eq. \ref{eq:FF_3D_front} we propose an estimate for the total width of gradient stabilized drainage fronts, we need to check whether the total extent of the tail front also satisfactorily scales as $|\nabla p|^{-\nu/(1+\nu)}$. For this, we simply calculate $\eta_{t}^*=z_{\eta t}^{max}-z_c$, where $z_{\eta t}^{max}$ is the maximum value of $z$ among the front tail bonds.

 Unlike the front tail calculated with Eq. \ref{eq:Gouyet_eta_t}, the results in Fig. \ref{fig:appendix_eta_f_p_pc} suggest that the total extent of the front tails, $\eta_{t}^*$, scales with $|\nabla p|^{-\nu/(1+\nu)}$ only in low and moderate values of $|\nabla p|$. For both investigated network topologies, when $|\nabla p|\gtrapprox0.075$, directly measured values of the front tail fall way below the theoretical prediction. For this reason, in the results presented in Sec. \ref{sec:results}, we suggest that our theoretical predictions of the front width are likely no longer valid above $|\nabla p|^{lim}\approx0.075$. 

 In the following Appendix section, we indicate how $|\nabla p|$ is estimated in our drainage IP simulations.

\section{Occupation gradient magnitude in gradient stabilized slow drainage in porous media}
\label{app:nabla_p_drainage}

In Appendix \ref{app:nabla_p}, we suggest that for high occupation probability gradients in gradient percolation, the scaling $|\nabla p|^{-\nu/(1+\nu)}$ for the total length of the infinite cluster front tail may fail. Thus, an equivalent $|\nabla p|$ for our drainage simulations should be defined, so that we can estimate an appropriate range of simulation parameters for which our theoretical predictions are expected to be valid. 

In the context of two-dimensional stable drainage,  Eq. \ref{eq:occ_prob} was proposed by \citet{birovljev1991gravity} to estimate the difference in invasion probability between two pore throats at the front separated by a distance $\Delta z$ in the direction of the gradient. 
If we consider this distance to be equal to one pore size, the difference in invasion probability is equivalent to $\nabla p \big|_{p_c}a$, and given by:

\begin{equation}
    \nabla p|_{p_c}a = \int_{P_{crit}}^{P_{crit}+Ga} N(P_t) \,dP_t 
    \label{eq:nabla_p}
\end{equation}

Approximating the solution of Eq. \ref{eq:nabla_p} by the lowest order term of the Taylor expansion, and considering that in the case of stabilizing gravitational effects, $G=-\Delta \rho g$, we get $\nabla p \big|_{p_c}a \approx -N(P_{crit})\Delta \rho ga$. As the variation in occupation probability in two-dimensional stable drainage fronts is narrow, we can consider this gradient to be approximately valid for the full front extent. However, 3D stable drainage fronts extend over a large range of $p$, meaning that variation in $\nabla p$ may occur, if $N(P_t)$ is non-uniform. Still, since we simply aim to define an approximation for $\nabla p$ to estimate a validity range for drainage parameters, we adopt $|\nabla p|a =N(P_{crit})\Delta \rho ga$. 

Using this approximation, we can associate the limit gradient of occupation probability found in Appendix \ref{app:nabla_p}, $|\nabla p|^{lim}\approx0.075$, which represents the difference in $p$ over one lattice unit, to the invasion probability difference over the length of a pore, leading to $0.075=N(P_{crit})\Delta \rho ga$. Therefore, we suggest that results from drainage simulations with a density difference between the phases $\Delta \rho>0.075/(N(P_{crit})ga)$ may not be comparable to the theoretical framework presented in Sec. \ref{sec:theo}. 

\section{Values of $\mathcal{C}$ for the 3D stable drainage front width estimate}
\label{app:const_c}

In Appendix \ref{app:nabla_p}, we investigate the effect of the magnitude of $\nabla p$ on the scaling of the front tail total extent, $\eta_{t}^*$, in gradient percolation. There, we empirically define a limit for $\nabla p$, below which our front tail measurements follow $\eta_{t}^* \propto |\nabla p|^{-\nu/(1+\nu)}$. This limit is relevant to establish a range of drainage parameters within which our stable drainage front width estimate (see Eq. \ref{eq:FF_3D_front}) may be applicable. 

To predict the front width, we take a step further and attempt to define a prefactor $\mathcal{C}$ that fulfills $\eta_{t}^* = \mathcal{C}|\nabla p|^{-\nu/(1+\nu)}$. While the front tail scaling, based on percolation theory, is independent of the network topology, the results presented in Fig. \ref{fig:appendix_eta_f_p_pc} suggest that $\mathcal{C}$ should depend on it. Comparing the results of Figs. \ref{fig:eta_f_p_pc_Diamond} and \ref{fig:eta_f_p_pc_SC}, we verify that the total extent of the front tail is larger using the diamond-cubic network than with the simple-cubic network. 

Therefore, using the gradient-percolation results in Fig. \ref{fig:appendix_eta_f_p_pc} to fit the theoretical-scaling curves -- shown as the green continuous lines -- we estimate $\mathcal{C}=0.90$ for the simple-cubic network and $\mathcal{C}=1.55$ for the diamond cubic lattice. This range of values is in agreement with the correlation length scaling in percolation theory $\xi=\xi_0|p-p_c|^{-\nu}$, where $\xi_0$ should be of the order of unity \cite{sapoval1985fractal}. Furthermore, these values of $\mathcal{C}$ lead to a reasonable match between theoretical and numerical front widths, as presented in Sec. \ref{sec:front_width}. Still, a more rigorous way of defining $\mathcal{C}$ for Eq. \ref{eq:FF_3D_front} should be established in future studies.

\bibliography{biblio}

\end{document}